\documentclass{article}
\usepackage{graphicx}

\usepackage{color}
\usepackage{mathtools}
\usepackage{float}
\usepackage{amssymb}
\usepackage{url}
\usepackage{hyperref}
\usepackage{graphicx}
\usepackage{multirow}
\usepackage{epstopdf}

\usepackage{subcaption} 

\begin{document}

\title{Comparison of Two Mean-Field Approaches to Modeling An Epidemic Spread}

\author{\textit{Viktoriya Petrakova},  \\
             Intitute of compuatational modelling SB RAS  \\
              Academgorodok st., 50/44, Krasnoyarsk, Russia\\
             Sobolev Institute of Mathematics SB RAS \\
             Ac. Koptyuga ave., 4, Novosibirsk, Russia\\
              vika-svetlakova@yandex.ru \\    
              \textit{Olga Krivorotko},  \\
              Sobolev Institute of Mathematics SB RAS \\
              Ac. Koptyuga ave., 4, Novosibirsk, Russia\\
              krivorotko.olya@mail.ru}

\maketitle
\begin{abstract}
The paper  describes and compares three approaches to modeling an epidemic spread. The first approach is a well-known system of SIR ordinary differential equations. The second is a mean-field model, in which an isolation strategy for each epidemiological group (Susceptible, Infected, and Removed) is chosen as an optimal control. The third is another mean-field model, in which isolation strategy is common for the entire population. The considered models have been compared analytically, their sensitivity analysis has been carried out and  their predictive capabilities have been estimated using sets of synthetic and real data. For one of the mean-field models, its finite-difference analogue has been devised. The models have also been assessed in terms of their applicability for predicting a viral epidemic spread.  
\end{abstract}


\section{Introduction}

The mean-field models the originally formulated by \cite{Lasry_2007,LionsLec2007,Huang2006,Huang2007}, are becoming an increasingly popular tool for mathematical modeling. The enormous advantage of the mean-field approach is that it allows one to describe the collective behavior of multiple agents (players) making strategic decisions with a small number of equations, which significantly reduces the calculation time and computational complexity. Mathematical epidemiology in this sense is no exception, since the mean-field approach makes it easy to describe the interaction of individuals in a certain population where a virus is spreading, so they have to  make strategic decisions about e.g. isolation. The Covid-19 pandemic has given a huge impetus to the development of the entire field of mathematical epidemiology,  as in terms of tools for predicting epidemic spreads, as in terms of   estimating the effectiveness of preventive measures and assessment of their impact on the socio-economic life of the population.

A broad overview of the mean-field models simulating coronavirus spread is given in the review article \cite{Roy_2023_EpidMFGOverview}. Typically, these epidemiological models describe the temporal dynamics of virus spread with compartmental SIR models (see a review of such models \cite{Krivorotko_Kabanikhin_overview}) and a population as agents interacting within a strategic situation, whose strategic choices are determined by the Markov processes. Apart from this general approach, there have been many formulations of mean field epidemiological models, e.g.,  in \cite{Doncel_2020}, the selfish vaccination problem in the SIR model is described by a mean-field game with a finite number of states in continuous time; or in \cite{Bremaud_2022}, a model accounting for age differentiation is considered by merging the mean-field and evolutionary games. There are many other common formulation (see \cite{Roy_2023_EpidMFGOverview}). Searching for  successful mathematical solutions  for epidemiological problems within the mean-field approach is an actual problem.

It is noteworthy, however, that the number of studies where the prediction results obtained using mean field models would be compared against real statistical data remains quite small. We think that this is due, firstly, to the need to determine the model parameters that describe the spread of the virus, which leads to solving additional problems (in particular, ill-posed inverse ones), as well as the problem of collecting relevant data. Secondly, the well-posedness of the formulation of mean field problems imposes a number of restrictions on the functions used in the model that  reduces their effectiveness in describing real situations.

Previously, in our works, we proposed two mathematical formulations of the mean field for predicting the spread of Covid-19 and compared them with real statistical data in Novosibirsk \cite{Petr_SIRC_22} and Krasnoyarsk \cite{Petrakova_IEEE2}. Both statements pursued their goal and qualitatively showed forecasts that were closer to real data compared to basic compartmental epidemiological models. However, a detailed description of the second, later model has never been presented, nor has there been an objective comparison of them, either with each other or with the basic differential model used to describe the dynamics of the virus within the mean field approach. However, such a comparison is one of the key and most frequently asked questions from the scientific community.

The present study was aimed to achieve several goals. First hand, it was designed to once again demonstrate one of the approaches to the formulation of mean-field problems, proposed by V. Shaidurov and V. Petrakova(Kornienko) in \cite{Shaydurov_2020,Shaydurov_2021} that allows one to avoid some restrictions on the cost function in the model by transition to a discrete formulation of the optimization problem that inherits the properties of a continuous one. Second, within this study we  considered two mean-field epidemiological models formulated similarly but conceptually different and provided a comprehensive comparison of them. Third, we tried to determine the similarities, differences, and the degree of dependence of the presented mean field models on the choice of the basic differential SIR model that clusters the population relative to the immune status of individuals.

The article is organized as follows. Section \ref{sec_formulation} presents the formulations of the differential models under study, describes the ideas that underlie them, demonstrates the key properties of the models, and raises questions of the existence and uniqueness of their solution. Section \ref{sec_finite_diff} presents the finite-difference formulation of mean-field epidemiological models and describes their properties. Section \ref{sec_sensativity} presents a comparison of the considered models regarding their sensitivity to the determination of their parameters and initial distributions. Section \ref{sec_numerical_comparision} presents a numerical study of the models depending on different cost function chosen, and presents their key similarities and differences identified in practice. And finally, in Section \ref{sec_example_real}, the models under consideration are used to predict the development of the real epidemiological situation in Novosibirsk in 2020.

\section{Mathematical formulations of models for comparison} \label{sec_formulation}
\subsection{SIR differential model} 
For the purposes of comparison, let us consider  the well-known compartmental SIR epidemiological model \cite{Kermack_27}, written as a system of ordinary differential equations:
\begin{equation}
	\label{eq_SIR}
	\left\{
	\begin{aligned}
		& {d m_S}\big/{dt} = -\beta m_S m_I,\\
		& {d m_I}\big/{dt} = \beta m_S m_I - \gamma m_I, \\
		& {d m_R}\big/{dt} = \gamma m_I. \\ 
	\end{aligned}
	\right.
\end{equation}
Here the entire population is divided into three groups: $m_S(t)$ is the proportion of susceptible individuals to the virus; $m_I(t)$ is the infected part of population; $m_R(t)$ -- those who have recovered or died due to virus. The probability of transition between one epidemiological group ($\beta,\gamma$) and the initial state of the population are used to separate groups at each time moment $t$.
\begin{equation}
	\label{eq_SIR_init}
	m_i(0)=m_{i0}.
\end{equation}
Note that model \eqref{eq_SIR},\eqref{eq_SIR_init} fulfills the law of mass balance: $m_S(t)+m_I(t)+m_R(t)=1\; \forall t$.

\subsection{SIR EGC MF model} \label{sec_MFC_EEGC}
For the second formulation, consider the mean-field model, proposed in \cite{Lee_21} and then developed in \cite{Petr_SIRC_22}. Here only the model’s mathematical formulation and its main properties are presented. For more detailed description, see \cite{Petr_SIRC_22}.

\textit{We find the minima of cost functional}
\begin{equation}
	\label{eq_cost_func}
	\begin{aligned}
		J_{EGC}({{m}_{SIR}},{{\alpha }_{SIR}})=\int_{0}^{T}&{\int_{0}^{1}{\left( \underset{i\in\{S,I,R\}}{\sum} G_i\left( {{m}_{SIR}},{{\alpha }_{i}} \right) m_i \right. +}}  \\
		&\left. +g\left( t,x,{{m}_{SIR}} \right) \right)\text{d}x\text{d}t+\int\limits_{0}^{1}{\Phi \left( {{m}_{SIR}}\left( T,x \right) \right)}\text{d}x   
	\end{aligned}
\end{equation}
\textit{with restrictions in the form of the system of convection-diffusion equations }
\begin{equation}
	\label{eq_MFG_EEGC}
	\left\{ \begin{aligned}
		& \begin{aligned}
			{{\partial }_{t}}{{m}_{S}}+\nabla \left( {{m}_{S}}{{\alpha }_{S}} \right)+ 
			\beta {{m}_{S}}{{m}_{I}}-\sigma _{S}^{2}\Delta {{m}_{S}}/2\ =0,  
		\end{aligned} \\ 
		& \begin{aligned}
			{{\partial }_{t}} {{m}_{I}}+\nabla \left( {{m}_{I}}{{\alpha }_{I}} \right)-\beta {{m}_{S}}{{m}_{I}}+\gamma {{m}_{I}}-\sigma _{I}^{2}\Delta {{m}_{I}}/2\ =0,  
		\end{aligned} \\ 
		& 
		\begin{aligned}
			{{\partial }_{t}}{{m}_{R}}+\nabla \left( {{m}_{R}}{{\alpha }_{R}} \right)-\gamma {{m}_{I}}-\sigma _{R}^{2}\Delta {{m}_{R}}/2\ =0        
		\end{aligned}
	\end{aligned} \right.
\end{equation}
\textit{with initial} 
\begin{equation}
	\label{eq_MFG_EEGC_init} 
	m_i(0,x) = m_{0i}(x) \text{ on } \Omega
\end{equation}
\textit{and Neumann boundary conditions} 
\begin{equation}
	\label{eq_MFG_EEGC_bound} 
	\partial m_i/ \partial x = 0\; \forall t \text{ and } x \in \Gamma_\Omega.  
\end{equation}
Here the stochastic processes within the population are described using non-negative parameters $\sigma_i$, $i\in \{S,I,R\}$ where   $m_i(t,x): [0,T]\times \Omega \rightarrow \mathrm{R}$  are the functions presenting the distribution of individuals in each epidemiological group $i\in \{S,I,R\}$ over the state space $\Omega$ at each time moment $t\in[0,T]$. State variable $x$ indicates the population's loyalty to quarantine measures; $x=0$ is the agent's dedication to the imposed restriction measures and $x=1$ is the opposite. Functions $\alpha_i(t,x):[0,T]\times[0,1]\rightarrow \mathrm{\textbf{R}}$, $i \in\{S,I,R\}$ denote the compliance strategy of the representative agent of each group in the population.  Note also that 
\begin{equation}
	\label{conserv_for_initial_time}
	\int_0^1 \underset{i\in \{S,I,R\}}{\sum} m_{0i}(x)\text{d}x = 1.
\end{equation}
In \eqref{eq_cost_func} $G_i$ is the running cost of strategy implementation by each epidemiological group in population; $g$ is the function of current expenses that doesn't depend on isolation strategy; and $\Phi$ is terminal cost function. Here and after the symbol $\cdot_{SIR}$ denotes that an expression is presented as a combination of values $\cdot_S,\cdot_I$ and $\cdot_R$.

To get the optimal condition, the Lagrange multiplier method was applied, so, the first equation in \eqref{eq_MFG_EEGC} was multiplied by smooth function $\psi_S(t, x)\in C ^ {\infty} \left(\left[0, T \right] \times [0,1] \right)$ and the resulting expression was integrated by parts with respect to $t$ and $x$:
\begin{equation}
	\begin{aligned}
		\label{eq_Ls}
		&{L_S^{EGC}}:=-\int_{0}^{T}{\int_{0}^{{1}}{\left( {\partial \psi_S}/{\partial t}\;+{{\sigma_S }^{2}}\Delta \psi_S/2+\alpha_S \cdot  \partial \psi_S/\partial x \right)m_S\,\text{d}x \,}\text{d}t} + \\
		&+\int_{0}^{T}{\int_{0}^{{1}}}(\beta {{m}_{S}}{{m}_{I}}\psi_S)\text{d}x\text{d}t \,+\\
		&+\int_{0 }^{{1}}{\left( \psi_S(T,x)m_S(T,x)-\psi_S(0,x){{m}_{S0}}(x) \right)}\text{ d}x =0. 
	\end{aligned}   
\end{equation}
The same was done to the remaining equations in the system \eqref{eq_MFG_EEGC} to obtain $L_I^{EGC},\;L_R^{EGC}$. Note that  \eqref{eq_Ls} and other similar expressions were performed when the following boundary conditions were satisfied:
\begin{equation}
	\label{eq_psi_bound}
	\partial \psi_i/ \partial x = 0\; \;\forall t \in [0,T] \;\text{ and } \;x = 0,1\; {\forall i \in \{S,I,R\}},  
\end{equation}
and 
\begin{equation}
	\label{eq_alpha_EEGC_bound}
	\alpha_i(t,0) = \alpha_i(t,1) = 0\; \;\forall t \in [0,T] \; {\forall i \in \{S,I,R\}}. 
\end{equation}
Now, the Lagrange function corresponding to the optimization problem \eqref{eq_cost_func}--\eqref{eq_MFG_EEGC_bound} under consideration can be written down as:
\begin{equation}
	\begin{aligned}
		\label{eq_EEGC_Lagrange}
		\Im_{EGC} (m_{SIR},\alpha_{SIR},\psi_{SIR}):=J_{EGC} (m_{SIR},\alpha_{SIR}) - L_S^{EGC} - L_I^{EGC} - L_R^{EGC}.
	\end{aligned}    
\end{equation}
Variation \eqref{eq_EEGC_Lagrange} with respect to components $m_i\; \forall (t,x)\in [0,T]\times[0,1]$, $\forall i\in\{S,I,R\}$  led to the conjugate system 
\begin{equation}
	\label{eq_EEGC_HJB}
	\hspace{-5mm}\left\{ \begin{aligned}
		& 
		\begin{aligned}
			{\partial \psi_S}/{\partial t}\;+{{\sigma_S }^{2}}\Delta \psi_S/2&+\alpha_S \cdot \partial \psi_S / \partial x + \beta m_I (\psi_I-\psi_S) =\\ 
			&=
			-\underset{i\in\{S,I,R\}}{\sum}{m_i\partial G_i}/{\partial m_S}-G_S\,-{\partial g}/{\partial m_S},  
		\end{aligned}
		\\ 
		&\begin{aligned}
			{\partial \psi_I}/{\partial t}\;+{{\sigma_I }^{2}}\Delta \psi_I/2&+\alpha_I \cdot \partial \psi_I / \partial x + \beta m_S (\psi_I-\psi_S) + \gamma (\psi_R-\psi_I)=\\ 
			&=
			-\underset{i\in\{S,I,R\}}{\sum}{m_i\partial G_i}/{\partial m_I}-G_I\,-{\partial g}/{\partial m_I},  
		\end{aligned}\\
		& 
		\begin{aligned}
			{\partial \psi_R}/{\partial t}\;+{{\sigma_R }^{2}}\Delta \psi_R/2&+\alpha_R \cdot \partial \psi_R / \partial x =\\ 
			&=
			-\underset{i\in\{S,I,R\}}{\sum}{m_i\partial G_i}/{\partial m_R}-G_R\,-{\partial g}/{\partial m_R}  
		\end{aligned}
		\\
	\end{aligned}
	\right.
\end{equation}
and the``initial'' condition resulting from the terminal cost
\begin{equation}
	\label{eq_EEGC_HJB_initial} 
	\psi_i(T,\,x)=\frac{\partial\Phi}{\partial m_i}(T,x)\,\,\,\,\,\forall \text{ }x\in \left[ 0,1 \right], \; \forall i\in\{S,I,R\}.
\end{equation}
Variation \eqref{eq_EEGC_Lagrange} with respect to components $\alpha_i\; \forall (t,x)\in [0,T]\times[0,1]$ produced
the following optimality conditions for $\bar{\alpha_i} \in \mathrm{R}$ in addition to system \eqref{eq_psi_bound}, \eqref{eq_EEGC_HJB}, \eqref{eq_EEGC_HJB_initial}
\begin{equation}
	\label{eq_EEGC_alpha_opt}  
	\frac{\partial G_j}{\partial \bar{\alpha_i }}+\frac{\partial \psi_i}{\partial x}=0
\end{equation}
$\forall\, i\in \{S,I,R\}\,\;\forall \left( t,x \right)\in [0,T]\times \left[ 0,1 \right]$. Obtained coupled PDE systems  \eqref{eq_MFG_EEGC}--\eqref{eq_MFG_EEGC_bound} and \eqref{eq_psi_bound}, \eqref{eq_EEGC_HJB}, \eqref{eq_EEGC_HJB_initial} together with system of algebraic equations \eqref{eq_EEGC_alpha_opt} give the necessary conditions for the minimization of \eqref{eq_cost_func}. For clarification and to avoid confusion, this epidemiological model has been named \textit{SIR Mean Field Model with Control in Each Group (SIR EGC MF model)}.

\subsection{SIR TGC MF model} \label{sec_MFC_TEGC}
The third model included in our comparative analysis is based on the assumption that an isolation strategy is identical for the entire population and does not differ for epidemiological groups. On space-time domain $[0,T]\times \Omega$  one finds function $\alpha (t,x)$ equivalent to $\alpha_S,\alpha_I,\alpha_R$  from the previous case. At the same time it does not depend on the epidemiological status of the representative player. This formulation has been named  \textit{SIR Mean Field Model with Total Control for All Epidemiological Groups (SIR TGC MF model)}. In this case, the system \eqref{eq_MFG_EEGC} is rewritten as
\begin{equation}
	\label{eq_MFG_TEGC}
	\left\{ \begin{aligned}
		& \begin{aligned}
			{{\partial }_{t}}{{m}_{S}}+\nabla \left( {{m}_{S}}{{\alpha }} \right)+ 
			\beta {{m}_{S}}{{m}_{I}}-\sigma _{S}^{2}\Delta {{m}_{S}}/2\ =0,  
		\end{aligned} \\ 
		& \begin{aligned}
			{{\partial }_{t}} {{m}_{I}}+\nabla \left( {{m}_{I}}{{\alpha }} \right)-\beta {{m}_{S}}{{m}_{I}}+\gamma {{m}_{I}}-\sigma _{I}^{2}\Delta {{m}_{I}}/2\ =0,  
		\end{aligned} \\ 
		& 
		\begin{aligned}
			{{\partial }_{t}}{{m}_{R}}+\nabla \left( {{m}_{R}}{{\alpha }} \right)-\gamma {{m}_{I}}-\sigma _{R}^{2}\Delta {{m}_{R}}/2\ =0       
		\end{aligned}
	\end{aligned} \right.
\end{equation}
with the same initial \eqref{eq_MFG_EEGC_init} and boundary conditions \eqref{eq_MFG_EEGC_bound}. Note that system \eqref{eq_MFG_TEGC} differs from \eqref{eq_MFG_EEGC} only in the generality of control for all epidemiological groups. In contrast to \eqref{eq_cost_func},  the cost function for this formulation is written as:
\begin{equation}
	\label{TEGC_cost_func}
	\begin{aligned}
		J_{TGC}({{m}_{SIR}},{{\alpha }})=\int_{0}^{T}&{\int_{0}^{1}{\left( \underset{i\in\{S,I,R\}}{\sum} G_i\left( {{m}_{SIR}},{{\alpha }} \right) m_i \right. +}}  \\
		&\left. +g\left( t,x,{{m}_{SIR}} \right) \right)\text{d}x\text{d}t+\int\limits_{0}^{1}{\Phi \left( {{m}_{SIR}}\left( T,x \right) \right)}\text{d}x.  
	\end{aligned}
\end{equation}
Using the Lagrange multiplier method and repeating steps \eqref{eq_Ls} relative to system \eqref{eq_MFG_TEGC} we obtained the Lagrange function corresponding to the SIR TGC MF optimization problem:
\begin{equation}
	\begin{aligned}
		\label{eq_TEGC_Lagrange}
		\Im_{TGC} (m_{SIR},\alpha,\psi_{SIR}):=J_{TGC} (m_{SIR},\alpha) - L^{TGC}_S - L^{TGC}_I - L^{TGC}_R.
	\end{aligned}    
\end{equation}
The expression \eqref{eq_TEGC_Lagrange} is valid if conditions \eqref{eq_psi_bound} and 
\begin{equation}
	\label{eq_alpha_TEGC_bound}
	\alpha(t,0) = \alpha(t,1) = 0\; \;\forall t \in [0,T] 
\end{equation}
are satisfied, so  the conjugate system of partial differential equations can be written in the following way:
\begin{equation}
	\label{eq_TEGC_HJB}
	\hspace{-5mm}\left\{ \begin{aligned}
		& 
		\begin{aligned}
			{\partial \psi_S}/{\partial t}\;+{{\sigma_S }^{2}}\Delta \psi_S/2&+\alpha \cdot \partial \psi_S / \partial x + \beta m_I (\psi_I-\psi_S) =\\ 
			&=
			-\underset{i\in\{S,I,R\}}{\sum}{m_i\partial G_i}/{\partial m_S}-G_S\,-{\partial g}/{\partial m_S},  
		\end{aligned}
		\\ 
		&\begin{aligned}
			{\partial \psi_I}/{\partial t}\;+{{\sigma_I }^{2}}\Delta \psi_I/2&+\alpha \cdot \partial \psi_I / \partial x + \beta m_S (\psi_I-\psi_S) + \gamma (\psi_R-\psi_I)=\\ 
			&=
			-\underset{i\in\{S,I,R\}}{\sum}{m_i\partial G_i}/{\partial m_I}-G_I\,-{\partial g}/{\partial m_I},  
		\end{aligned}\\
		& 
		\begin{aligned}
			{\partial \psi_R}/{\partial t}\;+{{\sigma_R }^{2}}\Delta \psi_R/2&+\alpha \cdot \partial \psi_R / \partial x =\\ 
			&=
			-\underset{i\in\{S,I,R\}}{\sum}{m_i\partial G_i}/{\partial m_R}-G_R\,-{\partial g}/{\partial m_R}
		\end{aligned}
		\\
	\end{aligned}
	\right.
\end{equation}
with the condition on time horizon \eqref{eq_EEGC_HJB_initial}. 

A variation of \eqref{eq_TEGC_Lagrange} in respect to $\alpha$ gives an optimal condition for $\bar{\alpha} \in \mathrm{R}$ $\forall (t,x)\in [0,T]\times[0,1]$
\begin{equation}
	\label{eq_TEGC_alpha_opt}  
	\underset{j\in\{S,I,R\}}{\sum}m_j\left(\frac{\partial G_j}{\partial \bar{\alpha }}+\frac{\partial \psi_j}{\partial x}\right)=0.
\end{equation}

The presented EGC and TGC formulations of formally differ not only in the use of control general for the entire population or individual for each epidemiological group, but also in the form of optimal conditions. Formally, conditions \eqref{eq_TEGC_alpha_opt} differ in the presence of a sum of all ``particular'' control. However, this entails differences in some features of the models, for example, reactions to terminal conditions, as it was shown in \cite{Petrakova_IEEE2}, and a smaller contribution of a particular epidemiological group in determining this strategy.

\subsection{Properties of the considered models}
\subsubsection{Differential (compartmental) SIR model}

One of the most important properties of the SIR model in question  is the introduction of the basic reproduction number \cite{Kermack_27} 
$$\mathfrak{R_0} = \frac{\beta}{\alpha},$$
characteristic of morbidity and epidemic spread, so when $\mathfrak{R_0}>1$, the system describes an uncontrolled outbreak.

The solution of \eqref{eq_SIR},\eqref{eq_SIR_init} is unique due to the uniform boundedness of the partial derivatives of the right-hand sides over $m_i,\; i\in \{S,I,R\}$. A more detailed analytical study of the SIR model is presented in \cite{Semendyaeva_22}.  The authors showed that despite its simplicity and conciseness, the SIR model demonstrates non-trivial analytical behavior. In particular, the ODE system has non-isolated equilibrium points. The coordinates of the equilibrium points depend on the model’s initial values and parameters  and indicatethe system’s sensitivity of the system to the determination of these values.

\subsubsection{SIR MF models} \label{sec_prop_MFC}
The proposed epidemiological mean-field models  are known as continuous dynamic one. Note that there are a number of other approaches for modeling the dynamics of epidemic spread, e.g., , Aurell at al. \cite{Aurell_Stackelberg} considers using the Stackelberg mean-field game model between the main and mean fields of agents whose states develop in a finite state space; and Doncel et al. \cite{Doncel_2020} analyze an average game model of SIR dynamics in which players choose when to vaccinate. Ullah et al. \cite{Ullah_MFGFrac} have proposed a fractional order mean field model for vaccination games in which an epidemic spread and individual decisions must evaluate social behavior. Tembine et al. \cite{Tembine_MFG} specifies a class of mean-field-type  games that have discrete-continuous state spaces. A recent study by Roy et al. \cite{Roy_2023_EpidMFGOverview} has delved into the relationship between mean-field games and mean-field control models in relation to epidemiology, as well as the differences between these types of models. It should be noted that the approach proposed in these works is not widely discussed, and since most of the works on epidemiological mean-field models are based on mean-field games with a finite number of states, the approach presented in this paper has almost never been investigated. 

Now, let us formulate the important properties of the models described in sections \ref{sec_MFC_EEGC} and \ref{sec_MFC_TEGC}.

\textbf{Property 1.} \textit{Local law of mass conservation for the SIR EGC MF model. }  Under conditions \eqref{eq_MFG_EEGC_bound},\eqref{conserv_for_initial_time},\eqref{eq_alpha_EEGC_bound} the following equality
\begin{equation}
	\label{eq_EEGC_mas_cov}
	\frac{\partial }{\partial t}\int\limits_{0}^{1}{\sum\limits_{i\in \{S,I,R\}}{{{m}_{i}}(t,x)}}dx=0    
\end{equation}
is satisfied. Physically it means that the total mass of the epidemiological groups is conserved $\forall t$.

\textbf{Property 2.} \textit{Local law of mass conservation for the SIR TGC MF model.} If conditions \eqref{eq_MFG_EEGC_bound},\eqref{conserv_for_initial_time},\eqref{eq_alpha_TEGC_bound} are satisfied, equality \eqref{eq_EEGC_mas_cov} is true for SIR TGC MF model  \eqref{eq_MFG_TEGC}.

Properties 1 and 2 stipulate that the Neumann conditions are necessary for the fulfillment of the law of mass conservation for the entire population, which is the main idea of basic SIR-type models.

\textbf{Property 3.} \textit{On the reachability of an extremum.}  In accordance with the Lagrange multiplier method, conditions \eqref{eq_psi_bound},\eqref{eq_EEGC_HJB}--\eqref{eq_EEGC_alpha_opt} for SIR EGC MF and \eqref{eq_psi_bound},\eqref{eq_alpha_TEGC_bound},\eqref{eq_TEGC_HJB}  for SIR TGC MF are not sufficient but necessary for an extremum solution to exist. The well-posedness of the formulation must be determined by the choice of the function and the analysis of the second derivatives of the corresponding Lagrangian with respect to the required variables. However, if the cost function that describes real processes doesn't have convexity properties, it can lead to an incorrect formulation of the problem. In practice, convergence to an extremum can also be estimated numerically by calculating the corresponding values of the cost function, but this method does not guarantee that the optimum is achieved over the entire functional space. Another significant limitation for the described systems follows from the fact that even if sufficient extremum conditions are met, the required quantities are, in the general case, only local extremum points. The global extremum can follow from the uniqueness of the solution to the conjugate system of the described type, however, as it will be discussed below, the uniqueness of the solution to such systems, in the general case, does not hold. Therefore, when talking about solving a conditional minimization problem, we mean finding a local minimum that is closest to the given initial population distributions.

The main problem of the mean-field game theory is the existence and uniqueness of solutions to such problems. Several works prove a solution to the problem exists in the form of a system of conjugate differential equations of the type considered. Moreover, proving the existence of a solution to the system is usually not very difficult, e.g., Gomes et al. \cite{Gomes_ExistStationaryMFG} have proven there is a solution to the conjugate system of Fokker-Planck and Hamilton-Jacobi-Bellman equations in the stationary case for a Hamiltonian system of certain type. Lasry and Lions who founded  the theory of mean-field games, have demonstrated \cite{Lasry_2007} that a solution to the adjoint system exists if the running, current and terminal costs are the Lipschitz continuous functions, and the system’s  Hamiltonian is convex with respect to the control (strategy) variable. Most of the special cases that prove the existence of a solution to a system of conjugate partial differential equations have been presented in the relatively recent monograph \cite{Achdou_MFG_monograph}.

The uniqueness of a solution to a problem is a more complex issue. Lasry and Lions have shown \cite{Lasry_2007} that if the costs are not only the Lipschitz functions, but also monotone, then the solution to the problem is unique. At the same time, examples are described in the literature \cite{Bardi_NonUniq}--\cite{Cecchin_NonUniq} when, if the monotonicity condition is violated, the solution to the conjugate problem is indeed not unique. In \cite{Bardi_NonUniq} is also shown the uniqueness holds for system if the time horizon is ``short'', but there is no notes on choosing an appropriate time horizon value. 

However, the works presented above prove the existence and uniqueness for models describing dynamics in the form of convection-diffusion equations with a zero right-hand side (Fokker-Planck equations). Conditions for the existence and uniqueness of the mean field models presented here can be obtained in a weak formulation from \cite{Poreta_2015} and \cite{Ataei_2023}, so the main results from \cite{Poreta_2015} can be reformulated for our case. 

\textbf{Property 4. } \textit{Conditions for the existence and uniqueness of a weak solution of the SIR EGC MF model.} Assume that running cost function $G_i$ grows quadratically with respect to control function $\alpha_i$ and the following conditions are satisfied: 
\begin{equation}   \label{existence_unique_conditions}
	\begin{aligned}
		&\exists M>0: \;\; \left(\frac{\partial G_i}{\partial \alpha_i} (m_{SIR}, \alpha_{SIR}) - \frac{\partial G_i}{\partial \alpha_i} (m_{SIR}, \alpha_{SIR}) \right) \cdot (\alpha_i - \tilde{\alpha_i})>0 \\     
		& \forall \alpha_i,\tilde{\alpha_i}:\alpha_i\neq\tilde{\alpha_i}, \forall i\in \{S,I,R\}
		\text{ whenever } \left| \frac{\partial G_i}{\partial \alpha_i}\right|, \; \left| \frac{\partial G_i}{\partial \tilde{\alpha}_i}\right|>M.
	\end{aligned}
\end{equation}
Assume that $g$, $\Phi$ are non decreasing with respect to $m$, bounded below and satisfy
\begin{equation}
	\begin{aligned}
		\label{existence_unique_conditions_2}
		&\forall L>0 \; g_L:=\underset{ m\in[0,L]}{\sup}\left| g(t,x,m)\right|\in L^1\left( (0,T)\times\Omega\right), \\
		&\forall L>0 \; \Phi_L:=\underset{ m\in[0,L]}{\sup}\left| \Phi(T,x)\right|\in L^1\left(\Omega\right), \\
		& f_i \in L_1 \left( (0,T)\times\Omega\right),\; \forall i \in \{S,I,R\},
	\end{aligned}
\end{equation}
where $f_i,\; i\in\{S,I,R\}$ are the right parts of \eqref{eq_MFG_EEGC}: 
\begin{equation}
	\label{eq_right_parts_of_EEGC} 
	f_S=-\beta m_S m_I; \; 
	f_I=\beta m_S m_I-\gamma m_I; \; 
	f_R=\gamma m_I. \\ 
\end{equation}
Then, for any $m_{0i}\in L^\infty(\Sigma)_+$ such as $\log m_{0i}\in L^1(\Omega)$, there exists unique weak solution $(m_{SIR},\psi_{SIR})$ to the system of \eqref{eq_cost_func}--\eqref{eq_MFG_EEGC_bound},\eqref{eq_psi_bound},\eqref{eq_alpha_EEGC_bound},\eqref{eq_EEGC_HJB},\eqref{eq_EEGC_HJB_initial}. 

\textbf{Property 5. } \textit{Conditions for the existence and uniqueness of a weak solution of the SIR TGC MF model.}
The same conditions \eqref{existence_unique_conditions},\eqref{existence_unique_conditions_2} are valid for SIR TGC MF, taking into account its features regarding the generality of control for the entire population.

\section{Finite-difference analogue of SIR MF models} \label{sec_finite_diff}

As it was shown in section \ref{sec_prop_MFC} despite the conceptual simplicity of the mean-field approach, the use of such models in the continuous statement follows together with the existence and uniqueness problems. Instead, we propose finite-difference analogue of conditional minimization problem (example.g., SIR EGC MF \eqref{eq_cost_func}--\eqref{eq_MFG_EEGC_bound}), which  is an independent problem and inherits the basic properties of the differential one.

To do so, let us introduce a grid that is uniform in time
\[t_{k} =k\tau ,\, \, \, \, \, k=0,...,M,\, \, \, \tau =T\big/ M\]
and a space
\[x_{i+1/2} =(i+1/2)h,\, \, \, \,\, i=-1,...,N,\, \, \, h={1/N} .\]
The agent distribution is considered as piecewise linear functions  $m^{i,h} (t,x)$ at each time level $t_{k}$, which are continuous at $[0,1]$ and linear in each segment $\omega _{j} =\left[x_{j-1/2} ,x_{j+1/2} \right]\,\,\forall \, j=1,...,N-1$. In addition, $m^{i,h} (t_{k} ,x)$ is assumed as constant at intervals $\omega _{0} =[0,x_{1/2}]$ and $\omega _{N} =\left[x_{N-1/2} ,1\right]$. The boundary conditions \eqref{eq_MFG_EEGC_bound} are replaced by  
\begin{equation}
	\label{eq_discr_bound_m}
	m_{k,-1/2}^{i,h}=m_{k,1/2}^{i,h}\text{ and } m_{k,N+1/2}^{i,h}=m_{k,N-1/2}^{i,h}.    
\end{equation}

\subsection{Finite-difference SIR TGC MF optimization problem} \label{sec_finite_diff_TGC}

Now let us consider a finite-difference analogue of the optimal control model \eqref{eq_TEGC_Lagrange}. Instead of \eqref{eq_MFG_TEGC} we will consider the following system of algebraic equations for each point $(t_k, x_{j+1/2})$, $\forall k = 1,...,M,\; \forall j = 0, ..., N-1$.  To do so, the Semi-Lagrange approximation proposed for SIRC MFG  model in \cite{Petr_SIRC_22} has been applied. It leads to the following finite-difference approximation of the equations from \eqref{eq_MFG_TEGC}:
\begin{equation}
	\label{eq_FPK_approxim}
	Dm_{k,j+1/2}^{i,h}=f_{k-1,j+1/2}^{i,h}+\Gamma m_{k-1,j+1/2}^{i,h}
\end{equation}
$\forall i\in \left\{ S,I,R \right\}$, where 
\begin{equation}
	\label{eq_FPK_DM_approxim}
	Dm_{k,j+1/2}^{i,h}=\left( \frac{1}{8\tau }-\frac{\sigma _{i}^{2}}{2{{h}^{2}}} \right)m_{k,j-1/2\;}^{i,h}+\left( \frac{3}{4\tau }+\frac{\sigma _{i}^{2}}{{{h}^{2}}} \right)m_{k,j+1/2\;}^{i,h}+\left( \frac{1}{8\tau }-\frac{\sigma _{i}^{2}}{2{{h}^{2}}} \right)m_{k,j+3/2\;}^{i,h},  
\end{equation}
\begin{equation}
	\label{eq_FPK_GAMMA_approxim}
	\Gamma m_{k-1,j+1/2}^{i}=\gamma _{k,j+1/2\;}^{i,1}m_{k-1,j-1/2\;}^{i,h}+\gamma _{k,j+1/2\;}^{i,2}m_{k-1,j+1/2\;}^{i,h}+\gamma _{k,j+1/2\;}^{i,3}m_{k-1,j+3/2\;}^{i,h} 
\end{equation}
with coefficients
\begin{equation}
	\begin{aligned}
		\label{eq_Gamma_coeff_TEGC}
		& \gamma _{k,j+1/2\;}^{i,1}=\frac{1}{8\tau }\left( 1+{4\tau \alpha _{k,j}}/{h}\; \right), \\ 
		& \gamma _{k,j+1/2\;}^{i,2}=\frac{1}{8\tau }\left( 3+{4\tau \alpha _{k,j}}/{h}\; \right)+\frac{1}{8\tau }\left( 3-{4\tau \alpha _{k,j+1}}/{h}\; \right), \\ 
		& \gamma _{k,j+1/2\;}^{i,3}=\frac{1}{8\tau }\left( 1-{4\tau \alpha _{k,j+1}}/{h}\; \right). \\ 
	\end{aligned}
\end{equation}
Notation $f_{\cdot ,\cdot }^{i,h}$ in \eqref{eq_FPK_approxim} represents a grid analogue of \eqref{eq_right_parts_of_EEGC}:
\begin{equation}
	\label{eq_FPK_rightSide_approxim} 
	\begin{aligned}
		& f_{k-1,j+1/2}^{S,h}=-\beta m_{k-1,j+1/2}^{S,h}m_{k-1,j+1/2}^{I,h}, \\ 
		& f_{k-1,j+1/2}^{I,h}=\beta m_{k-1,j+1/2}^{S,h}m_{k-1,j+1/2}^{I,h}-\gamma m_{k-1,j+1/2}^{I,h}, \\ 
		& f_{k-1,j+1/2}^{R,h}=\gamma m_{k-1,j+1/2}^{I,h}. \\ 
	\end{aligned}
\end{equation}
The initial conditions corresponding to \eqref{eq_MFG_EEGC_init} for \eqref{eq_FPK_approxim}--\eqref{eq_FPK_rightSide_approxim} are added  in the discreate case:
\begin{equation}
	\label{eq_discr_initial_FPK}
	m_{0,j+1/2\;}^{i,h}=m_{0i}^{{}}({{x}_{j+1/2\;}})\;\forall j=0,...,N-1.
\end{equation}
To obtain a discrete optimization problem, the integral cost function in \eqref{TEGC_cost_func} is replaced by the discrete one: 
\begin{equation}
	\label{eq_discr_TEGC_cost_func}
\begin{aligned}
	{{J}^{h}_{TGC}}(m_{{}}^{SIR,h},\alpha _{{}}^{h})=&\tau h\sum\limits_{k=0}^{M}{\sum\limits_{j=0}^{N-1}{\left( \underset{i\in\{S,I,R\}}{\sum} r_{k,j+1/2\;}^{i,h}m^{h,i}_{k,j+1/2}+g_{k,j+1/2\;}^{h} \right)}}\\
	&+h\sum\limits_{j=0}^{N-1}{\Phi \left( m_{M,j+1/2}^{SIR,h} \right)}.
\end{aligned}
	\end{equation}
Here $r_{k,j+1/2\;}^{i,h}$ is carried out for $G_i\left( {{m}_{SIR}},{\alpha }\right)$ in the following way:
\begin{equation}
	\label{eq_discr_TEGC_r_func}
	r_{k,j+1/2\;}^{i,h}=G_i\left( m_{k,j+1/2\;}^{SIR,h},\alpha _{k,j}^h\right)/2\ +G_i\left( m_{k,j+1/2}^{SIR,h},\alpha _{k,j+1}^{h}\right)/2\   
\end{equation}
with $\alpha _{k,j}^{h}:={{\alpha }^{h}}\left( {{t}_{k}},{{x}_{j}} \right).$

Now we can formulate a discrete SIR TGC MF optimization problem: \textit{minimize the function \eqref{eq_discr_TEGC_cost_func} with a set of conditions in the form of algebraic restriction \eqref{eq_discr_bound_m}--\eqref{eq_discr_initial_FPK}}.

\textbf{Remark 1.} \textit{On the correspondence of a finite-difference problem to the corresponding continuous formulation.} The Taylor series expansion demonstrates that the expressions \eqref{eq_discr_bound_m}--\eqref{eq_discr_TEGC_r_func} approximate the corresponding differential ones \eqref{eq_MFG_EEGC_init}, \eqref{eq_MFG_EEGC_bound}, \eqref{eq_MFG_TEGC}, \eqref{TEGC_cost_func}  at each point $(t_k,x_{j+1/2})$ $\forall k =1,..., M;\; \forall i = 0,..., N-1$ with the order $O(\tau+h^2)$.

\textbf{Remark 2.} \textit{Monotonicity of approximation.} Let us impose the following conditions:
\begin{equation}
	\label{eq_restr_to_grid}
	{{h}^{2}}\le 4\tau \underset{i}{\mathop{\min }}\,\{\sigma _{i}^{2}\}\text{ and }\text{ }\tau |\alpha _{k,j}|\le h/8\text{ }\forall k=0,...M,\text{ }\forall j=0,...N-1.   
\end{equation}
Conditions \eqref{eq_restr_to_grid} guarantee that all $\gamma^i_{k,j+1/2}$ are positive, and \eqref{eq_discr_bound_m}--\eqref{eq_discr_initial_FPK} is monotonic approximation for \eqref{eq_MFG_EEGC_init}, \eqref{eq_MFG_EEGC_bound}, \eqref{eq_MFG_TEGC}.  It means that for non-negative initial values $m_{0,\cdot }^{i,h}$ $\forall i\in \left\{ S,I,R \right\}$ components $m_{k,\cdot }^{i,h}$ $\forall k\in \left\{ 1,...,M \right\}$ are non-negative too.  Proving this remark can be avoided, since the proof repeats the one for ``Proposition 1''  in \cite{Petr_SIRC_22}, and the differences between the current model and the proposed one in \cite{Petr_SIRC_22} do not critically affect its course.

\textbf{Remark 3.} \textit{Discrete analogue of the conservation law for a total mass of agents.} Let us sum the expressions in \eqref{eq_FPK_approxim} over $j= 0, ..., N-1$ and multiply the obtained result by $\tau h$.  For non-negative $m^{i,h}(t_{k-1},x)$ the following equality is satisfied: 
\begin{equation}
	\label{eq_discr_mass_coverg}
	\begin{aligned}
		\int\limits_{0}^{1}&{\sum\limits_{i=S,I,R}{{{m}^{i,h}}}({{t}_{k}},x)\text{d}x}=\int\limits_{0}^{1}{{{m}^{h}}({{t}_{k}},x)\text{d}x}=\\
		&=\int\limits_{0}^{1}{\sum\limits_{i=S,I,R}{{{m}^{i,h}}}({{t}_{k-1}},x)\text{d}x}=\int\limits_{0}^{1}{{{m}^{h}}({{t}_{k-1}},x)\text{d}x}.
	\end{aligned}
\end{equation}
since the following properties are performed: 
\begin{equation}
	\label{eq_gamma_coef_propert}
	\gamma _{k,j+3/2\;}^{i,1}+\gamma _{k,j+1/2\;}^{i,2}+\gamma _{k,j-1/2\;}^{i,3}=1/\tau \;\;\forall i\in \{S,I,R\}    
\end{equation}
and $\sum\limits_{i=S,I,R}{f_{k,j+1/2}^{i,h}}=0$ . Here $m^h$ is grid value of total mass of population.

\textbf{Remark 4.} \textit{Stability assessment}.  For \eqref{eq_FPK_approxim}--\eqref{eq_FPK_rightSide_approxim} with initial \eqref{eq_discr_initial_FPK} and boundary \eqref{eq_discr_bound_m} conditions the following assessment is performed $\forall i\in \{S,I,R\}$
\begin{equation}
	\label{eq_discr_stability_asses_FPK}
	\underset{0\le k\le M}{\mathop{\max }}\,{{\left\| {{m}^{i,h}}\left( {{t}_{k}},\cdot  \right) \right\|}_{1,h}}\le {{\left\| {{m}^i_{0}}\left( \cdot  \right) \right\|}_{1,h}}+T\underset{0\le k\le M}{\mathop{\max }}\,{{\left\| {{f }^{i,h}}\left( {{t}_{k}},\cdot  \right) \right\|}_{1,h}},	
\end{equation}
where ${{\left. \left\| {{m}^{i,h}}\left( {{t}_{k}},\cdot  \right) \right. \right\|}_{1,h}}$ is discrete analogue of $L_{1}(0,1)$–norm for the grid function.

\textbf{Proof.} The proof of the remark repeats that for  ``Proposition 2'' in \cite{Petr_SIRC_22}. The proof is based on summing up \eqref{eq_FPK_approxim} over $j=0,...,N-1$ for non-negative $f^{i,h}_{k-1,j+1/2}$.  after multiplying \eqref{eq_FPK_approxim} by $\tau$ and $h$. The key property of model coefficients \eqref{eq_gamma_coef_propert} entails assessment \eqref{eq_discr_stability_asses_FPK} for each time layer $k$.

\textbf{Remark 5.} \textit{About discrete optimization problem}. The obtained independent discrete minimization problem \eqref{eq_FPK_approxim} -- \eqref{eq_discr_TEGC_r_func} approximates a continuous optimization problem with order   and inherits its main property being the fulfillment of the population conservation law at each time level. Note that the ``independence'' of the formulation leads the solutions to the discrete and continuous optimization problems that may not generally coincide due to the approximation type of function. However, the proof of the existence of a solution to the discrete optimization problem is limited to the fact that the matrix of system \eqref{eq_FPK_approxim} is non-singular and complies with the restrictions on grid step \eqref{eq_restr_to_grid} and some non-burdensome restrictions on functionality to be discussed below.

To formulate a discrete conjugate problem for grid optimization, let us introduce a grid-function set $\psi _{\cdot ,\cdot }^{i,h}=\left\{ \psi _{k,j+\text{1}/\text{2}\;}^{i,h} \right\}_{j=0,...,N-1}^{k=0,...,M}.$ Multipling the $k-$th component of $i-$th equation from \eqref{eq_FPK_approxim} by $\psi _{k,\cdot }^{i,h}$ and sum over $k$, one obtains the discreate analogue of $L_{{}}^{i,h}$ in \eqref{eq_Ls} for SIR TGC MF model. Now let us write down a saddle point problem for the grid optimization one \eqref{eq_FPK_approxim} -- \eqref{eq_discr_TEGC_r_func}
\begin{equation}
	\label{eq_discr_TEGC_saddle_point}
	\underset{(m_{{}}^{SIR,h},\alpha _{{}}^{h})}{\mathop{\inf }}\,\underset{\psi _{{}}^{SIR,h}}{\mathop{\sup }}{{\Im }^{h}_{TGC}}(m_{{}}^{SIR,h},\alpha _{{}}^{h},\psi _{{}}^{SIR,h}):={{J}^{h}}(m_{{}}^{SIR,h},\alpha _{{}}^{h})-\sum\limits_{i=\{S,I,R\}}{{{L}^{i,h}}}.\text{ }
\end{equation}
The Lagrangian is differentiated  in \eqref{eq_discr_TEGC_saddle_point} with respect to the individual components to obtain the following system of algebraic equations for $i\in \left\{ S,I,R \right\}$:
\begin{equation}
	\label{eq_discr_HJB_approx}
	D\psi _{k,j+1/2}^{i,h}=z_{k+1,j+1/2}^{i,h}+\Gamma \psi _{k+1,j+1/2}^{i,h},
\end{equation}
\begin{equation}
	\label{eq_discr_GAMMA_HJB}
	\Gamma \psi _{k+1,j+1/2}^{i,h}=\gamma _{k+1,j-1/2}^{i,3\;}\psi _{k+1,j-1/2}^{i,h\;}+\gamma _{k+1,j+1/2}^{i,2\;}\psi _{k+1,j+1/2}^{i,h\;}+\gamma _{k+1,j+3/2}^{i,1\;}\psi _{k+1,j+3/2}^{i,h\;}
\end{equation}
and 
\begin{equation}
	\label{eq_discr_rightSide_HJB}
	\begin{aligned}
		& z_{k+1,j+1/2}^{S,h}=\beta m_{k+1,j+1/2}^{I,h}\left( \psi _{k+1,j+1/2}^{I,h}-\psi _{k+1,j+1/2}^{S,h} \right)\\
		&+r_{k+1,j+1/2}^{S,h}+\underset{i\in\{S,I,R\}}{\sum}{m_{k+1,j+1/2}^{i,h}}B_{k+1,j+1/2}^{i,S,h}+b_{k+1,j+1/2}^{S,h}, \\ 
		& z_{k+1,j+1/2}^{I,h}=\beta m_{k+1,j+1/2}^{S,h}\left( \psi _{k+1,j+1/2}^{I,h}-\psi _{k+1,j+1/2}^{S,h} \right)+\gamma \left( \psi _{k+1,j+1/2}^{R,h}-\psi _{k+1,j+1/2}^{I,h} \right)+ \\
		& +r_{k+1,j+1/2}^{I,h}+\underset{i\in\{S,I,R\}}{\sum}{m_{k+1,j+1/2}^{i,h}}B_{k+1,j+1/2}^{i,I,h}+b_{k+1,j+1/2}^{I,h},\\
		& z_{k+1,j+1/2}^{R,h}=r_{k+1,j+1/2}^{R,h}+\underset{i\in\{S,I,R\}}{\sum}{m_{k+1,j+1/2}^{i,h}}B_{k+1,j+1/2}^{i,R,h}+b_{k+1,j+1/2}^{R,h}. \\ 
	\end{aligned}    
\end{equation}

In \eqref{eq_discr_rightSide_HJB} $\gamma^{i,h}_{k,j+1/2}$ are determined by \eqref{eq_Gamma_coeff_TEGC}; $B_{{}}^{i,l,h}$ is presented in the approximation for $\displaystyle\frac{\partial G_i}{\partial m_l}_{{}}$ functions as:
\[B_{k,j+1/2\;}^{i,l,h}=\frac{\partial G_i}{\partial m_l}
\left( m_{k,j+1/2\;}^{SIR,h},\alpha _{k,j}^h\right)\Big/2\ +\frac{\partial G_i}{\partial m_l}\left( m_{k,j+1/2}^{SIR,h},\alpha _{k,j+1}^{h}\right)\Big/2,\]
$i,l\in\{S,I,R\}$ and \[b_{k,j+1/2}^{i,h}=\frac{{\partial g}}{\partial m_i}\left( {{t}_{k}},{{x}_{i+1/2}},m_{k,j+1/2}^{SIR,h} \right).\]

System \eqref{eq_discr_HJB_approx}--\eqref{eq_discr_rightSide_HJB} is supplemented with  the terminal conditions 
\begin{equation}
	\label{eq_discr_HJB_initial}
	D\psi _{i,h}^{M,j+1/2}=\partial {{\Phi }_{h}}/\partial m_{i,h}^{M,j+1/2} \;\forall i\in \{S,I,R\},\ \forall j\in \{0,N-1\}
\end{equation}
and boundary discrete conditions in analogue of \eqref{eq_discr_bound_m}:
\begin{equation}
	\label{eq_discr_HJB_bound}
	\psi_{k,-1/2}^{i,h}=\psi_{k,1/2}^{i,h}\text{ and } \psi_{k,N+1/2}^{i,h}=\psi_{k,N-1/2}^{i,h} \; \forall k \in \{0,...,M\}    
\end{equation}
$\forall \; i \in \{S,I,R\}.$ Note that for the discrete problem, the ``initial'' condition of discrete HJB equation also assume a one-time solution of the system of algebraic equations.

\textbf{Remark 6} \textit{(stability assessment for the conjugate problem).} For \eqref{eq_discr_HJB_approx}-- \eqref{eq_discr_HJB_bound} under the restrictions \eqref{eq_restr_to_grid} the following assessments are performed for $i\in \{S,I,R\}$:
\[\underset{0\le k\le M}{\mathop{\max }}\,{{\left\| {{\psi }^{i,h}}\left( {{t}_{k}},\cdot  \right) \right\|}_{\infty ,h}}\le {{\left\| {{\psi }^{i}}({{t}_{M}},\cdot ) \right\|}_{\infty ,h}}+T\underset{0\le k\le M}{\mathop{\max }}\,{{\left\| {{z}^{i,h}}\left( {{t}_{k}},\cdot  \right) \right\|}_{\infty ,h}},\]
where ${{\left\| {{\psi }^{i,h}}\left( {{t}_{k}},\cdot  \right) \right\|}_{\infty ,h}}$ is a discrete analogue of $L_{\infty}(0,1)$–norm for grid function.

\textbf{Proof.} Let us consider $\vert \Tilde \psi^{i,h}(t_k, x_{j+1/2}) \vert$ as the component reaching its maximum absolute value on the layer $t_k$ so that $\vert \Tilde \psi^{i,h}(t_k, x_{j+1/2}) \vert =\| \psi^{h,i}(t_k,\cdot) \|_{\infty,h}$. To obtain the required inequality use  the key property of the coefficients \eqref{eq_gamma_coef_propert} 
\[\| \psi^{h,i}(t_k,\cdot) \|_{\infty,h} = \vert \Tilde \psi^{i,h}(t_k, x_{j+1/2}) \vert
\leq  \| \psi^{h,i}(t_{k+1},\cdot) \|_{\infty,h} +\ \tau \| z^{h,i}(t_k,\cdot) \|_{\infty,h}\] again.
Mathematical induction on $k$ leads to
\begin{equation}
	\notag
	\| \psi^{h,i}(t_k,\cdot) \|_{\infty,h} 
	\leq  \| \psi^{h,i}(t_{M},\cdot) \|_{\infty,h} +\ (M-k)\tau \| z^{h,i}(t_k,\cdot) \|_{\infty,h}.  
\end{equation}
Taking a maximum over $k$ we obtain the required assessment.

By varying the discrete Lagrangian with respect to components $\alpha^h_{k,j}\in R$, we obtain the grid optimality conditions written as 
\begin{equation}
	\label{eq_discr_TEGC_alpha_opt} 
	\begin{aligned}
		&\underset{i\in\{S,I,R\}}{\sum} \frac{m^{i,h}_{k,j+1/2} + m^{i,h}_{k,j-1/2}}{2} 
		\left( \frac{\partial G_i}{\partial {\alpha^h_{k,j} }} +  \frac{\psi _{k,j+1/2\;}^{i,h}-\psi _{k,j-1/2\;}^{i,h}}{h} \right) = 0 
	\end{aligned}
\end{equation}
$\forall k=0,\ldots ,M\ \forall j=1,\ldots ,N-1$. 

\textbf{Remark 7} \textit{(about restrictions on functions).} Let's impose some restrictions on the functions to ensure the solvability and uniqueness of the system \eqref{eq_discr_TEGC_alpha_opt} :
\begin{equation}
	\begin{aligned}
		& \frac{\partial G_i}{\partial {\bar \alpha}} \text{ are continious and strictly monotonus for all admissible } \bar\alpha \in R; \\
		&  \frac{\partial G_i}{\partial {\bar \alpha}} \Big|_{x=0,1} = 0.
	\end{aligned}
\end{equation}
Therefore, a set of equation systems s  \eqref{eq_discr_bound_m} --\eqref{eq_discr_TEGC_r_func}, \eqref{eq_discr_HJB_approx}--\eqref{eq_discr_TEGC_alpha_opt} describes a discrete SIR TGC MF optimization problem

\subsection{Finite-difference SIR EGC MF optimization problem  }

We will not describe in detail the discrete formulation of the discrete SIR EGC MF optimization problem as it was done in paper \cite{Petr_SIRC_22} for a similar model with proof of the corresponding statements. Here we will only briefly indicate the main changes relative to the model described in Section \ref{sec_finite_diff_TGC} to understand the main differences between models.

Firstly, the cost function is rewritten as
\begin{equation}
	\label{eq_discr_EEGC_cost_func}
	\begin{aligned}
			{{J}^{h}_{GC}}(m_{{}}^{SIR,h},\alpha ^{SIR,h})=&\tau h\sum\limits_{k=0}^{M}{\sum\limits_{j=0}^{N-1}{\left( \underset{i\in\{S,I,R\}}{\sum} r_{k,j+1/2\;}^{i,h}m^{i,h}_{k,j+1/2}+g_{k,j+1/2\;}^{h} \right)}} \\
			&+h\sum\limits_{j=0}^{N-1}{\Phi \left( m_{M,j+1/2}^{SIR,h} \right)}, 
		\end{aligned}
\end{equation}
where $r_{k,j+1/2\;}^{i,h}$ now depends not only on one discrete control but on the controls in each epidemic group:
\begin{equation}
	\label{eq_discr_EEGC_r_func}
	r_{k,j+1/2\;}^{i,h}=G_i\left( m_{k,j+1/2\;}^{SIR,h},\alpha _{k,j}^{i,h}\right)/2\ +G_i\left( m_{k,j+1/2}^{SIR,h},\alpha _{k,j+1}^{i,h}\right)/2.\   
\end{equation}

Secondly, the discrete analogue of the convection-diffusion system for discrete SIR EGC MF problem is described by the same system \eqref{eq_discr_bound_m},\eqref{eq_FPK_approxim},\eqref{eq_discr_initial_FPK} but with other expressions for $\gamma^{i,h}_{k,+1/2}$:
\begin{equation}
	\begin{aligned}
		\label{eq_Gamma_coeff_EEGC}
		& \gamma _{k,j+1/2\;}^{i,1}=\frac{1}{8\tau }\left( 1+{4\tau \alpha^{i,h}_{k,j}}/{h}\; \right), \\ 
		& \gamma _{k,j+1/2\;}^{i,2}=\frac{1}{8\tau }\left( 3+{4\tau \alpha^{i,h}_{k,j}}/{h}\; \right)+\frac{1}{8\tau }\left( 3-{4\tau \alpha^{i,h}_{k,j+1}}/{h}\; \right), \\ 
		& \gamma _{k,j+1/2\;}^{i,3}=\frac{1}{8\tau }\left( 1-{4\tau \alpha^{i,h}_{k,j+1}}/{h}\; \right). \\ 
	\end{aligned}
\end{equation}

The corresponding conjugate system can be formulated for a system of algebraic equations \eqref{eq_discr_HJB_approx}, \eqref{eq_discr_HJB_initial}, \eqref{eq_discr_HJB_bound} taking into account \eqref{eq_discr_EEGC_cost_func} -- \eqref{eq_Gamma_coeff_EEGC}.

And finally, the key difference between models is reached when discrete optimal condition
\begin{equation}
	\label{eq_discr_EEGC_alpha_opt} 
	\frac{\partial G_i}{\partial {\alpha^{i,h}_{k,j} }} +  \frac{\psi _{k,j+1/2\;}^{i,h}-\psi _{k,j-1/2\;}^{i,h}}{h}  = 0\end{equation}
$\forall\; i \in \{S,I,R\}\;\forall k=0,\ldots ,M\ \forall j=1,\ldots ,N-1$. 
is satisfied.

The key properties, estimates  and constraints described in Remarks 1--6, remain valid when considering the SIR EGC MF model.

\section{Sensitivity analysis} \label{sec_sensativity}

As a rule, epidemiological models are presented as systems of differential equations. The constants used in the systems usually reflect the details of the modeled epidemiological processes in the region under consideration and must be obtained based on information about morbidity statistics, the nature of the virus and other comparable data. However, almost always their exact determination is not possible because the statistical data is inaccurate or noisy \cite{Krivorotko_2020}. Various factors, such as comorbidities, age differentiation, virus contagiousness, and population density, influence the speed at which the virus spreads. Therefore, the sensitivity of the constructed epidemiological model remains a significant indicator of its quality.

One method for assessing the global sensitivity of models is the Extended Fourier Amplitude Sensitivity Test (eFAST)\cite{Saltelli_1999}. Like more famous Sobol method \cite{Sobol_1990}, eFAST allows to divide the total variance of model output into components,
corresponding to the model’s input parameters.
The variance caused by any given parameter and their interaction is quantified by sensitivity indicesbeing measurable indicators of the model's sensitivity to parameter identification. The idea of global sensitivity methods consists in multiple calculations of model outputs when input parameters vary, i.e., in creating a sample of calculations from which the contribution of the variation of each parameter to the total variance will be assessed.
Compared to other approaches, eFAST requires a small sample size, making it a much more attractive method for estimating complex models.

For a general description of the approach, assume that the model whose sensitivity is analyzed can be represented as
\begin{equation}\label{eq_model_black_box}
	Y = f(q_1,\ldots, q_m),
\end{equation}
where $\vec q = (q_1,\ldots , q_m)^T$ is the vector of the model’s input parameters; $\vec Y = (y_1,...,y_n)^T$ is the vector of the model’s outputs. The first-order influence of parameter $q_i$ on the output is determined as:
\begin{equation}\label{eq_Sobol_Si}
	SI_i = \frac{V_{q_i}(\mathbb{E}_{\textbf{Q}_{i}}(Y|q_i))}{V(Y)},\quad i=1,\ldots, m.
\end{equation}
Here $\textbf{Q}_{i}$ is a matrix of dimension $N_s \times m-1$ with the rows of “pseudo-random” values of unknown parameters $\{q_1,...,q_{i-1},q_{i+1},q_m\}$ within specified boundaries. The detailed description of method can be found in \cite{Saltelli_1999,Petr_Sens_23} The larger the $SI_i$ value the more sensitive the $q_i$ parameter to the chosen output.

For brevity sake, we will not describe in detail the form in which the integrals responsible for calculating the sensitivity indices are written since it has been done in \cite{Petr_Sens_23} for the SEIR-HCD TGC MF model. To implement eFAST, we have used a ready-made implementation presented in the \textrm{SALib} module  (\url{https://salib.readthedocs.io/en/latest/index.html}) of the \textrm{Python} language.

The minimum required number of iterations $N_s$ with different input parameters to obtain sensitivity indices has been determined in accordance with the Nyquist theorem:
\begin{equation}
	\label{eq_Nyquist_theor}
	N_s = 2M_p \underset{i=1,...,n}{\max}\{w_i\}+1,
\end{equation}
where $M_p$ is the calculation parameter (number of harmonics) usually chosen as 4 or 6; $w_i$ is the frequency associated with each input along which output fluctuations are monitored. Note that frequencies $w_i,\; i =1,...,n$ must be chosen in such a way for  amplitude $pw_i\; p<M_p$ is a linear combination of the others. For some $n$ satisfying this condition, the corresponding sets of frequencies are written in \cite{Schaibly_1973}. Therefore, \eqref{eq_Nyquist_theor} determines the minimum required calculations for different sets of input parameters. For our calculations, we have chosen $M_p$ = 6, and $N_s= 24000$, which exceeds the minimum required number of calculations.

\subsection{Sensitivity of differential SIR model}

\begin{figure}[h]
	\centering
	\includegraphics[width=0.9\linewidth]{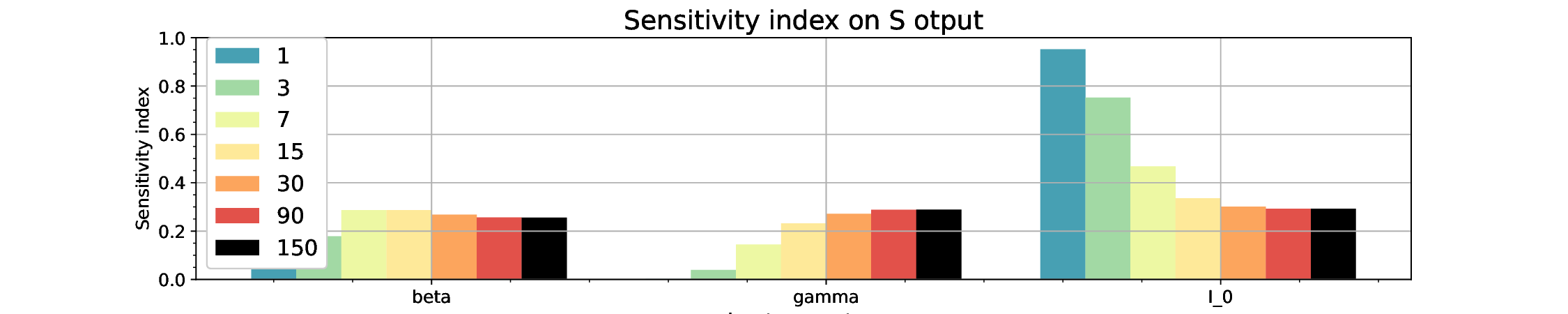}  \\
	\includegraphics[width=0.9\linewidth]{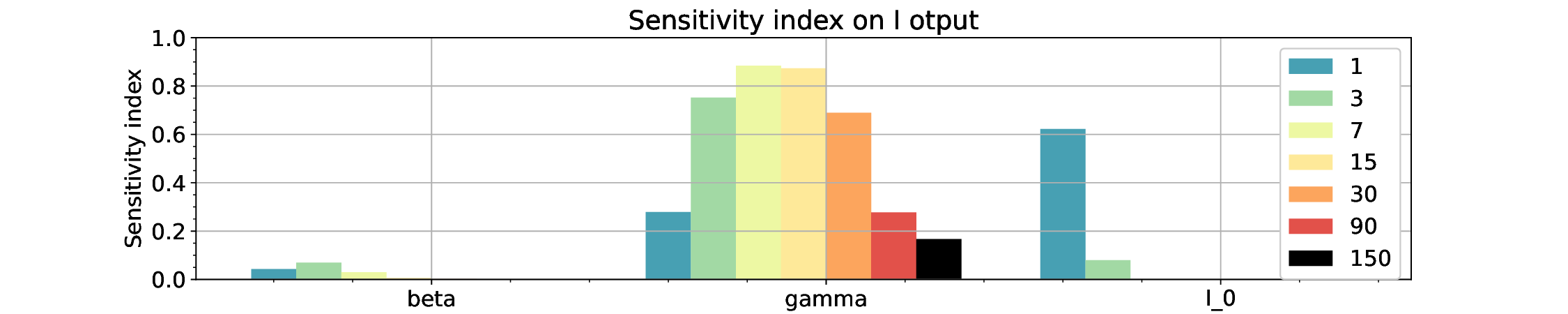} \\
	\includegraphics[width=0.9\linewidth]{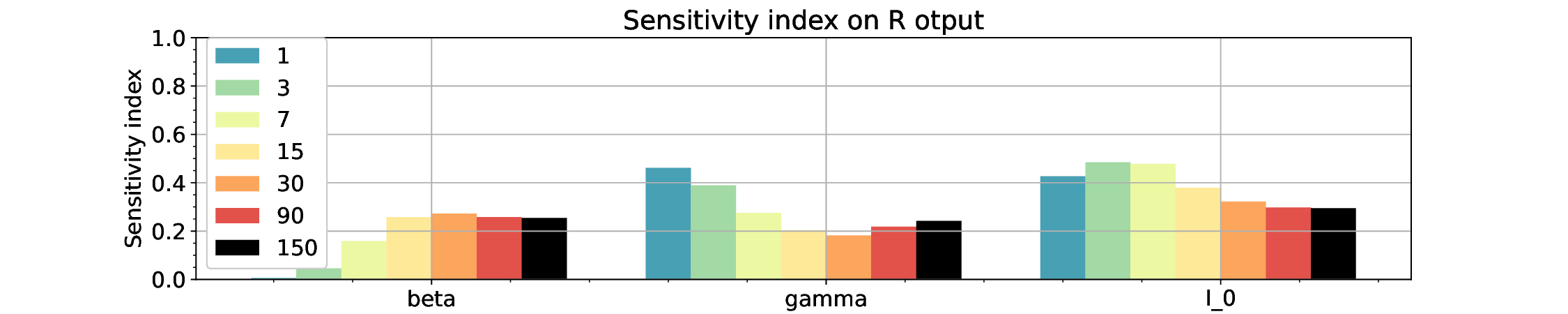}  \\
	\caption{Sensitivity indices of the differential model \eqref{eq_SIR},\eqref{eq_SIR_init} with a set of input parameters \eqref{SIR_input_parameters} for different simulation times: $T=1,3,7,15,30,90,150$ days }
	\label{fig_Sens_SIR_diff}
\end{figure}

To understand the time dynamics of the sensitivity of each of the models described in this paper, a differential SIR model has also been included in the study. As a set of input sensitivity parameters of the model, let us consider the vector of input parameters:
\begin{equation}
	\label{SIR_input_parameters}
	\vec q = (q_1,q_2,q_3) = (\beta,\gamma, I_0),
\end{equation}
where $I_0=m_I(0)$. As the output, the number of people in each epidemiological group at the time horizon $T$ is considered:
\begin{equation}
	\label{SIR_output}
	Y = (Y_S, Y_I, Y_R) = (m_S(T),m_I(T), m_R(T)).
\end{equation}
Here, the sensitivity of model \eqref{eq_model_black_box} presented in the form \eqref{eq_SIR},\eqref{eq_SIR_init} is evaluated for different $T$, where $T = 1,3,7,15,30,90,150$ days. Additionally, 
\[ m_S(0) = 1-I_0;\;\; m_I(0)=I_0;\;\; m_R(0)=0. \]
for initial data for the studied model, and the and the input parameters change within the following ranges: 
\[ q_1 \in [0,1];\;\; q_2 \in [0,1];\;\; q_3 \in [0.05,0.8]. \].
In figure \ref{fig_Sens_SIR_diff}, the values of $SI_i,\; i\in \{S,I,R\}$ are presented for the formulation mentioned above. 

For the chosen input parameters, the extremely sensitive for the SIR model are the initial value of the number of infected people $I_0$, as well as the $\gamma$ parameter, which is responsible for the recovery rate. In this case, the significance of $\gamma$  reaches its peak in the medium-term modeling period and decreases during long-term modeling. The sensitivity of the outputs to the value of $\beta$ parameter is an almost constant value and does not depend on the time period over which the simulation is performed.

\subsection{Sensitivity of SIR mean field models}

Sensitivity analysis of mean-field models is a much more difficult problem if compared to that for simple differential models. First, in addition to the epidemiological parameters inherited from the simple model, mean-field models are also characterized by stochastic parameters ($\sigma_S, \sigma_I, \sigma_R$). Secondly, the research in \cite{Petr_SIRC_22} showed that the final result of the modeling is significantly dependent on the choice of the initial distribution of parts of the population within each epidemiological group. At the same time, the study \cite{Petr_SIRC_22} have not revealed how significant this influence is and how it is determined. Here we will try to more strictly formulate the criterion and a set of values that require more strict identification. To achieve the stated goal for mean field models, we will assume that initial distributions of agents for SIR MF models for masses ${{m}_{i}}(0,x)$, $i\in\{S,I,R\}$, $x\in\Omega$ are determined by the following expressions:
\begin{equation}
	\label{eq_initial_Gauss_distr} 
	{{m}_{0i}}=\frac{{{A}_{i}}}{{{B}_{i}}}\left(
	\exp\left( -\frac{{{(x-x_{i}^{c})}^{2}}}{2{{(\sigma_{i}^{c})}^{2}}}
	\right)/\sigma_{i}^{c}\sqrt{2\pi
	}+{{a}_{i}}{{x}^{2}}+{{b}_{i}}{{(1-x)}^{2}} \right),
\end{equation}
where $A_i$ is the proportion of the current group in relation  to the total population at the initial time; $B_i$ is the normalization coefficient equal to the integral over $\Omega$ of the expression in brackets; $a_i = \exp\left(-{(1-x_i^c)^2}\Big/{2 (\sigma^c_i)^2}\right) (1-x^c_i)/(2(\sigma^ c_i)^3\sqrt{2\pi})$ and $b_i = \exp\left(-{(x_i^c)^2}\Big/{2 (\sigma^c_i)^2}\right) (x^c_i)/(2(\sigma^c_i)^3\sqrt{2\pi})$ ensure boundary conditions \eqref{eq_MFG_EEGC_bound} for $m(0,x)$. Physically, this means that we define the initial distributions as Gaussian (though with some small correction to comply with the boundary conditions), which is very convenient for research, since the normal distribution is determined by two parameters such as the average value that also determines the mode, i.e. peak mass of agents and the dispersion showing how much the peak value is spread across the state space. Therefore, to analyze the sensitivity of mean-field models, we have included parameters $x^c_i,\sigma^c_i$ $\forall i \in \{S,I,R\}$ in the set of input parameters of model \eqref{eq_model_black_box}, with respect to which the study is carried out. It should also be noted that the values of $A_i$ coincide with the corresponding initial values $m(0)$ of the simple differential model of SIR \eqref{eq_SIR},\eqref{eq_SIR_init}.

\begin{figure}[h]
	\centering
	\includegraphics[width=\linewidth]{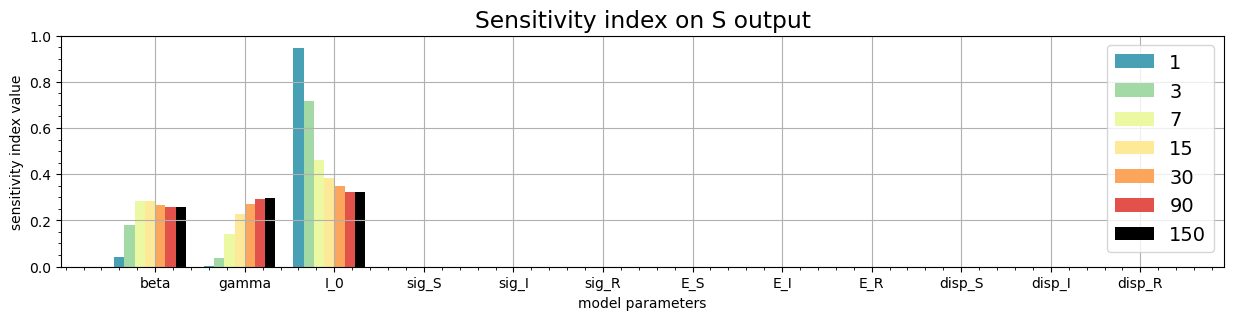} \\
	\includegraphics[width=\linewidth]{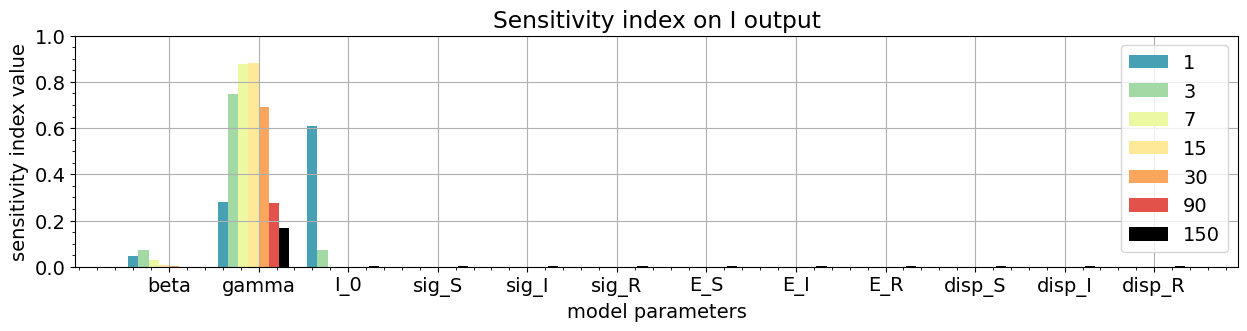} \\
	\includegraphics[width=\linewidth]{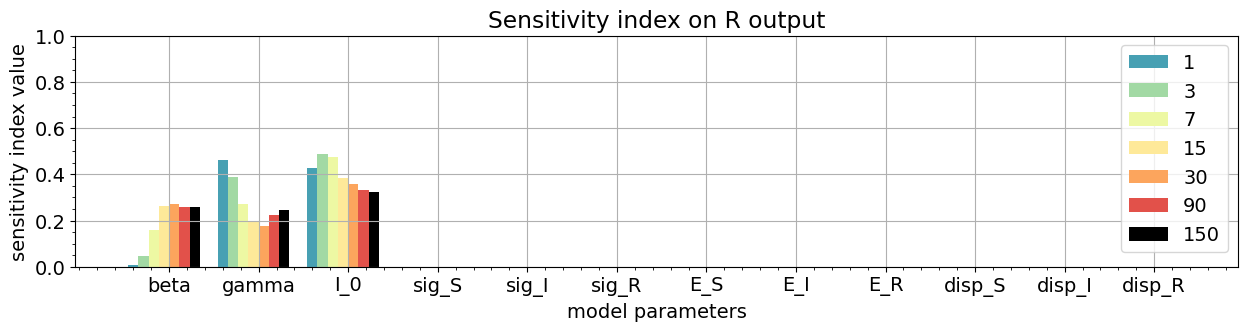}  \\
	\caption{Sensitivity indices of the SIR TGC MF model \eqref{eq_TEGC_Lagrange} with a set of input parameters \eqref{eq_MFC_total_input_parameters} for different simulation times: $T=1,3,7,15,30,90,150$ days }
	\label{fig_Sens_SIR_TGC}
\end{figure}

Now let us examine the mean field models for sensitivity with respect to the vector of input parameters:
\begin{equation}
	\label{eq_MFC_total_input_parameters}
	\begin{aligned}
		q = (\beta,\gamma, I_0, \sigma_S, \sigma_I, \sigma_R, x^c_S, x^c_I, x^c_R, \sigma^c_S, \sigma^c_I, \sigma^c_R).   
	\end{aligned}
\end{equation}
To save the output meaning of s for MF SIR models we will consider
\begin{equation}
	\label{eq_MFC_SIR_output}
	Y = (Y_S, Y_I, Y_R),\; \text{where } Y_i = \underset{\Omega}{\int} m(T,x) \text{d}x,\; \forall i\; \in\{S,I,R\}.   
\end{equation}
The figure \ref{fig_Sens_SIR_TGC} shows the resulting values of sensitive indices for the SIR TGC MF model. The result of the sensitive analysis for SIR EGC MF model is quite close to this one. The inscriptions in the figures correspond to the order in which parameters are written, defined in \eqref{eq_MFC_SIR_output}, so \textit{'sig\_S','sig\_I','sig\_R'} correspond to model parameters $\sigma_S,\sigma_I,\sigma_R$;  \textit{'E\_S','E\_I','E\_R'} for $x^c_S,x^c_I, x^c_R$ and \textit{'disp\_S','disp\_I','disp\_R'} for $\sigma^c_S,\sigma^c_I, \sigma^c_R$.

\begin{figure}[h]
	\centering
	\includegraphics[width=\linewidth]{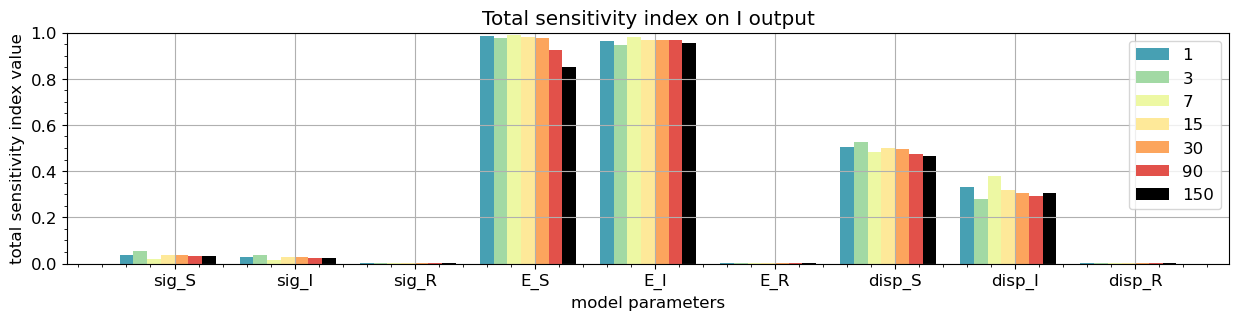} \\
	\caption{Total sensitivity indices of the SIR TGC MF model \eqref{eq_TEGC_Lagrange} for $I$ output for different simulation times: $T=1,3,7,15,30,90,150$ days }
	\label{fig__Total_Sens_SIR_TGC}
\end{figure}

From Figure \ref{fig_Sens_SIR_TGC}, it seems that the choice of initial distributions and stochastic parameters does not affect the simulation result. However, the numerical experiments demonstrated in \cite{Petr_SIRC_22} show exactly the opposite. This can be justified in the following way: the figure shows the sensitivity of each individual parameter relative to others, i.e., the contribution of each parameter to the total variance of the output. Firstly, it is clear that in this case the influence of epidemiological parameters is significantly greater than the variation of the initial distributions. Secondly, most likely, the contribution to the result is made not by the variation of a single parameter of the initial distribution (or stochastic parameter), but by their combination, which the first-order sensitivity index does not reflect. To test this hypothesis, consider the total sensitivity index $ST_i$. It examines the variance in the output, which encompasses all the variance resulting from the interaction of any input parameter's order with other parameters. The figure \ref{fig__Total_Sens_SIR_TGC}, present the total sensitivity indices $ST_i$ for the TGC model for a set of parameters $q = (\sigma_S, \sigma_I, \sigma_R, x^c_S, x^c_I, x^c_R, \sigma^c_S, \sigma^c_I, \sigma^c_R)$ for output \text{I}, which shows that the effect on output is determined by a combination of initial distributions.

\section{Numerical analysis of similarities and differences between models EGC and TGC MF approaches} \label{sec_numerical_comparision}

In this section, we examine the impact of various components of mean-field models (running, current, and terminal costs) on the simulation results Here that we are solving a discrete optimization problem, whose formulation is determined by the approach outlined in Section\ref{sec_finite_diff}. However, for convenience, we write the corresponding functionals in continuous form, implying that discretization \eqref{eq_discr_EEGC_cost_func}, \eqref{eq_discr_TEGC_cost_func} and restrictions \eqref{eq_FPK_approxim} are used for them. In other words, let us consider SIR EGC MF optimization problems  with the following functionals:

\begin{equation}
	\label{eq_EGC_quad_func}
	\begin{aligned}
		&J_{EGC}^{1}({{m}_{SIR}},{{\alpha_{SIR} }})=\int_{0}^{T}\int_{0}^{1}
		\underset{i\in\{S,I,R\}}{\sum}\frac{\alpha_i^2 m_i}{2} \text{d}x\text{d}t;
	\end{aligned}
\end{equation}

\begin{equation}
	\label{eq_EEGC_quad_current_func}
	\begin{aligned}
		&J_{EGC}^{2}({{m}_{SIR}},{{\alpha_{SIR} }})=\int_{0}^{T}\int_{0}^{1}
		\underset{i\in\{S,I,R\}}{\sum}\frac{\alpha_i^2 m_i}{2} + d_1 \left( m_I^2 + \left(1-m_S\right)^2\right) \text{d}x\text{d}t ;
	\end{aligned}
\end{equation}

\begin{equation}
	\label{eq_EEGC_quad_terminal_func}
	\begin{aligned}
		&J_{EGC}^{3}({{m}_{SIR}},{{\alpha_{SIR} }})=\int_{0}^{T}\int_{0}^{1}
		\underset{i\in\{S,I,R\}}{\sum}\frac{\alpha_i^2 m_i}{2} \text{d}x\text{d}t + d_2\int_0^1 m_I^2(T,x) \text{d}x;
	\end{aligned}
\end{equation}

\begin{equation}
	\label{eq_EGC_running_current_terminal_func}
	\begin{aligned}
		J_{EGC}^{4}({{m}_{SIR}},{{\alpha_{SIR} }})=\int_{0}^{T}\int_{0}^{1}
		\underset{i\in\{S,I,R\}}{\sum}\frac{\alpha_i^2 m_i}{2} +& d_1 \left( m_I^2 + \left(1-m_S\right)^2\right) \text{d}x\text{d}t \\
		&+ d_2\int_0^1 m_I^2(T,x) \text{d}x.
	\end{aligned}
\end{equation}

\begin{figure}[H]
	\centering
	\subcaptionbox{S group}{\includegraphics[width=0.33\textwidth]{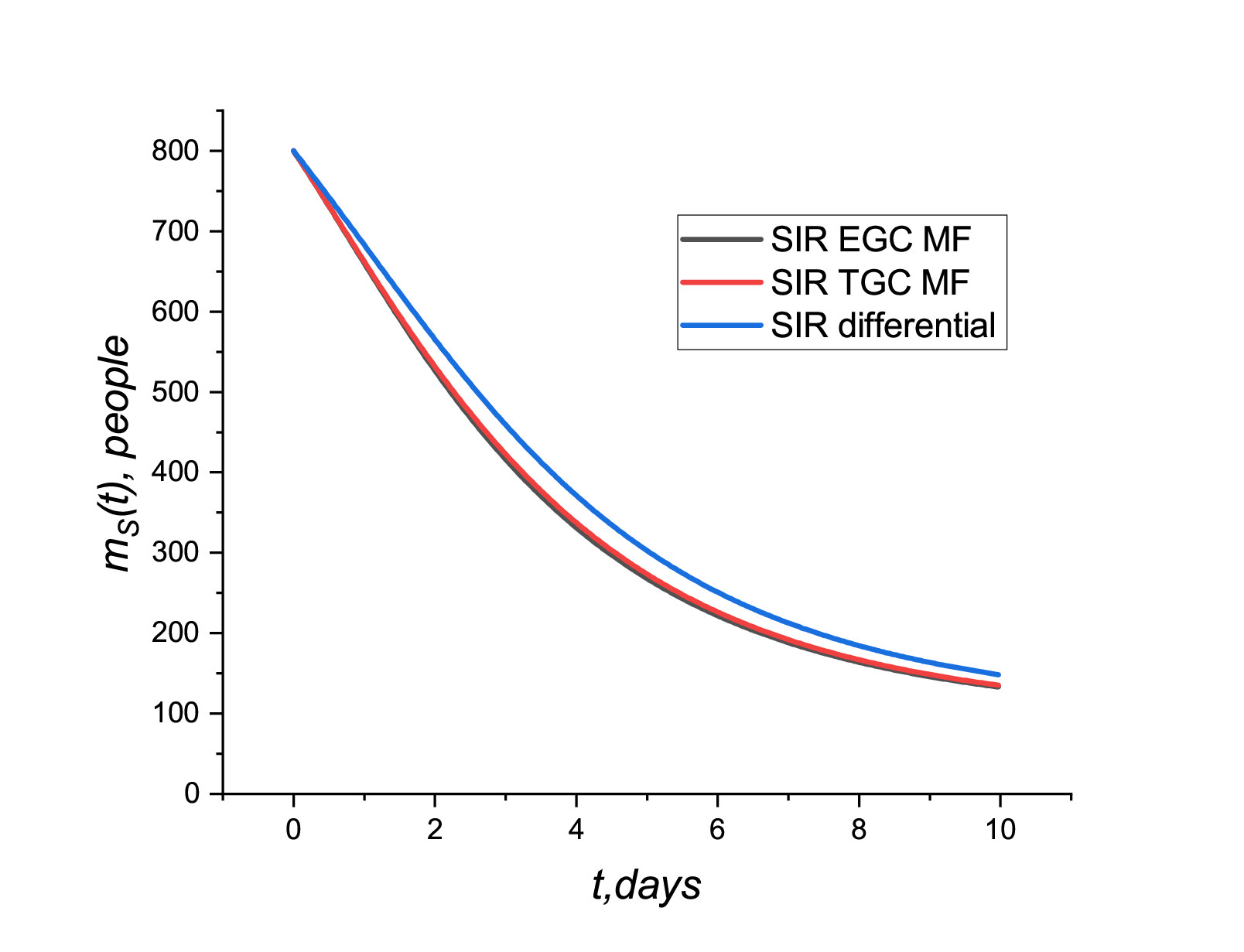}}%
	\hfill 
	\subcaptionbox{I group}{\includegraphics[width=0.33\textwidth]{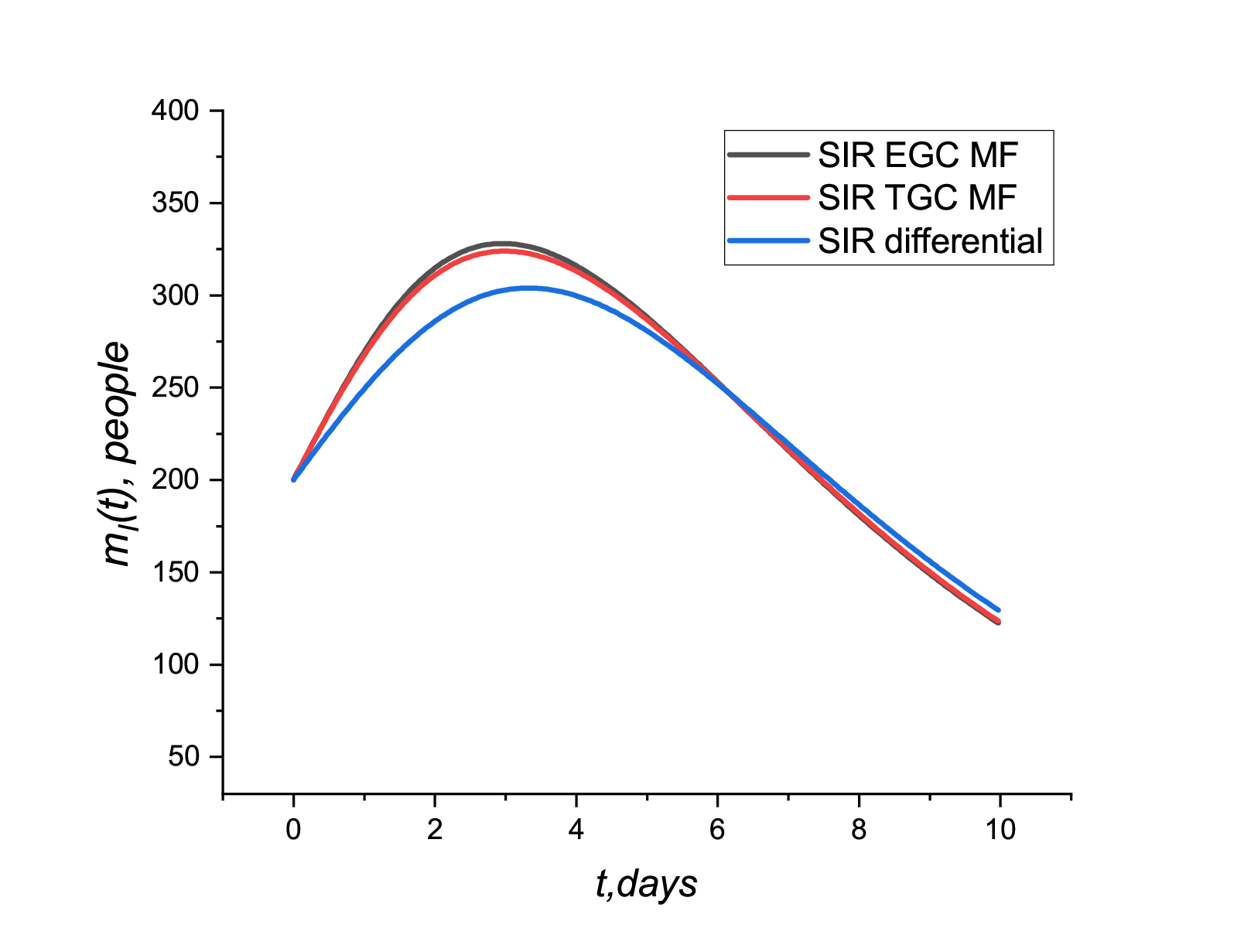}}%
	\hfill 
	\subcaptionbox{I group}{\includegraphics[width=0.33\textwidth]{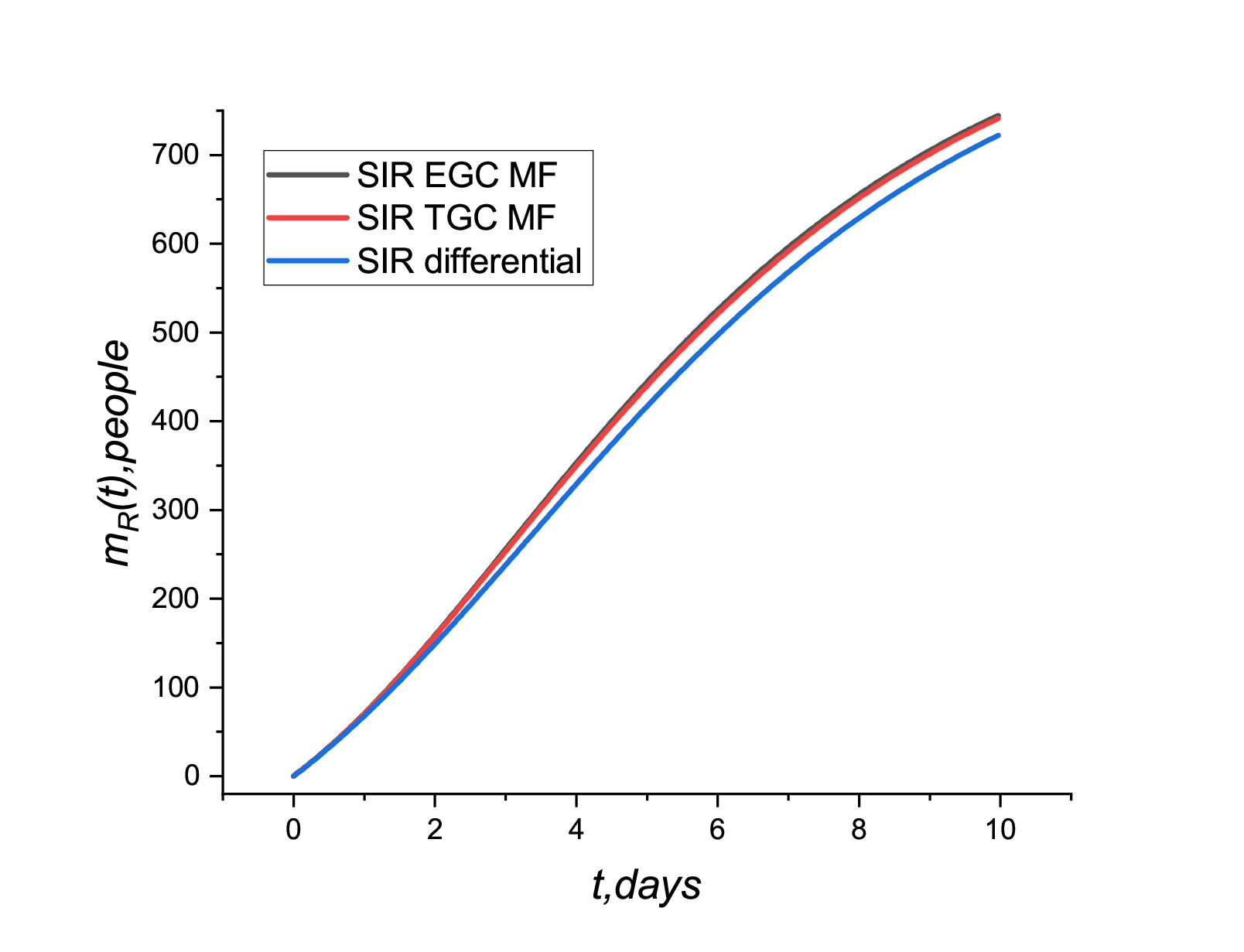}}%
	\caption{Comparison of simulation results of differential SIR model and mean field models with cost functionals $J_{EGC}^{3}$ and $J_{TGC}^{3}$}
	\label{fig_m_for_only_TerminalCost}
\end{figure}

The corresponding functionals $J_{TGC}^{q}$, $q=1,..,4$ for the SIR TGC MF model are obtained by replacing $\alpha_i$ in \eqref{eq_EGC_quad_func}--\eqref{eq_EGC_running_current_terminal_func}  with $\alpha$ which is general for entire population. Let us describe the general parameters used to run a computational experiment. Put
\begin{equation}
\notag
\begin{aligned}
T=10,\;&\beta=0.7,\; \gamma=0.3,\; S_0 = 0.8,\; I_0=0.2,\; R_0 = 0,\;\sigma_i=0.02,\\
&x_i^c = 0.5,\; \sigma_i^c = 0.1,\; d_1 = 0.002,\; d_2 = 10
\end{aligned}
\end{equation}
$\forall i\in\{S,I,R\}.$ The parameters have been chosen based on the following considerations. Epidemiological parameters $\beta,\gamma$ and the initial number of individuals in each population $S_0,I_0,R_0$ were chosen to ensure approximately comparable dynamics of each population group over a period of $T=$10 days. The development of the epidemiological situation under such parameters, obtained using the SIR model, is shown in Figure \ref{fig_m_for_only_TerminalCost}. The stochastic parameter $\sigma_i$ and the parameters of the initial distribution $x_i^c,\sigma_i^c$ were chosen small to minimize the influence of the diffusion term on the mean field systems. And parameters $d_1,d_2$ are responsible for balancing current and terminal costs in a numerical value. Note that we also compare the simulation results of both the mean-field and differential SIR models with the same parameters.

The mean-field models determined by functionals $J_{EGC}^{1}$ and $J_{TGC}^{1}$ have produced the same modeling result similar to that for he differential SIR model, since zero strategy is optimal for the EGC and TGC models, and the incidence dynamics is determined only by parameters $\beta$ and $\gamma$. 

\begin{figure}[H]
	\centering
	\subcaptionbox{$\alpha$ for TGC model}{\includegraphics[width=0.33\linewidth]{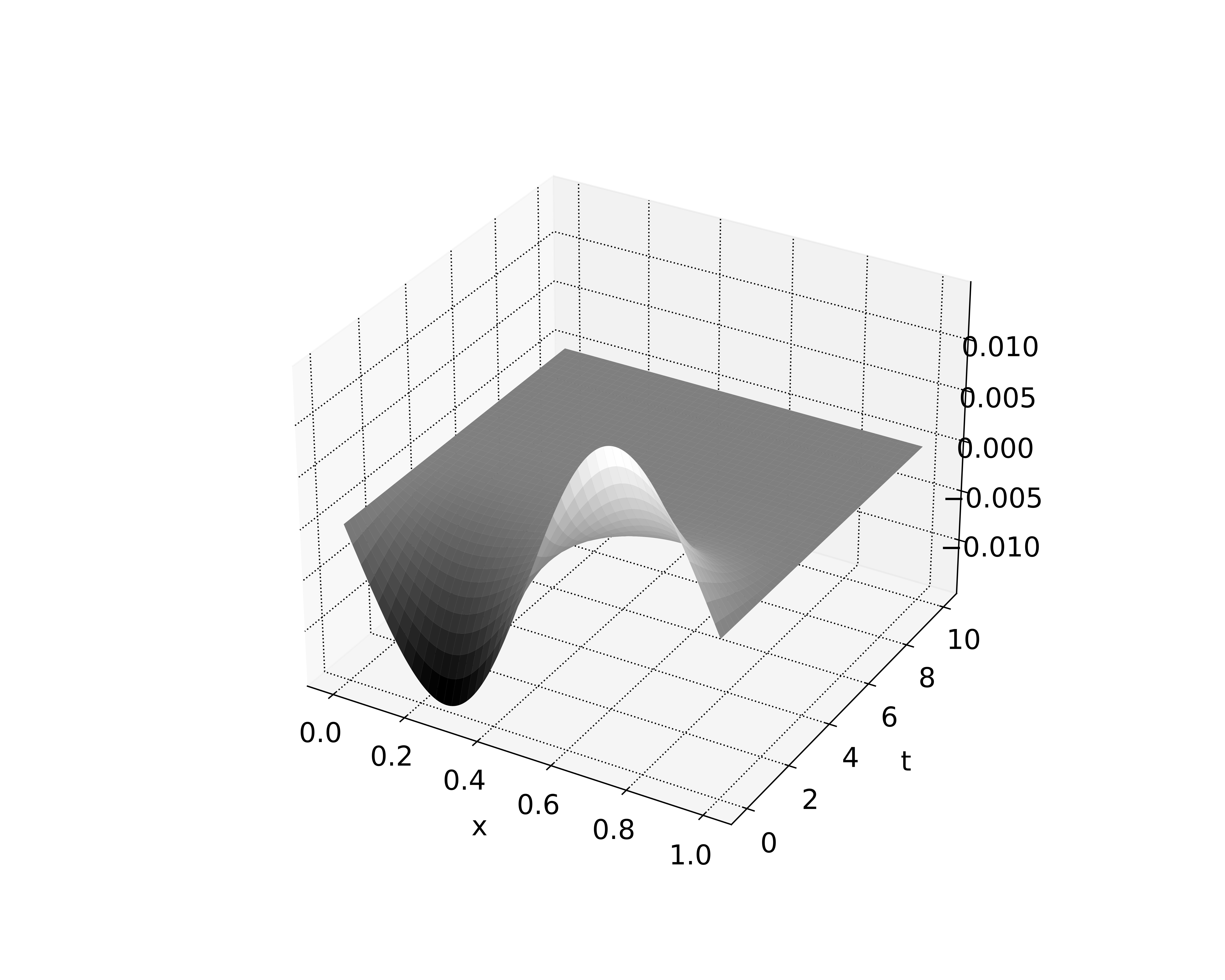}}%
	\hfill 
	\subcaptionbox{$\alpha_S$ for EGC model}{\includegraphics[width=0.33\linewidth]{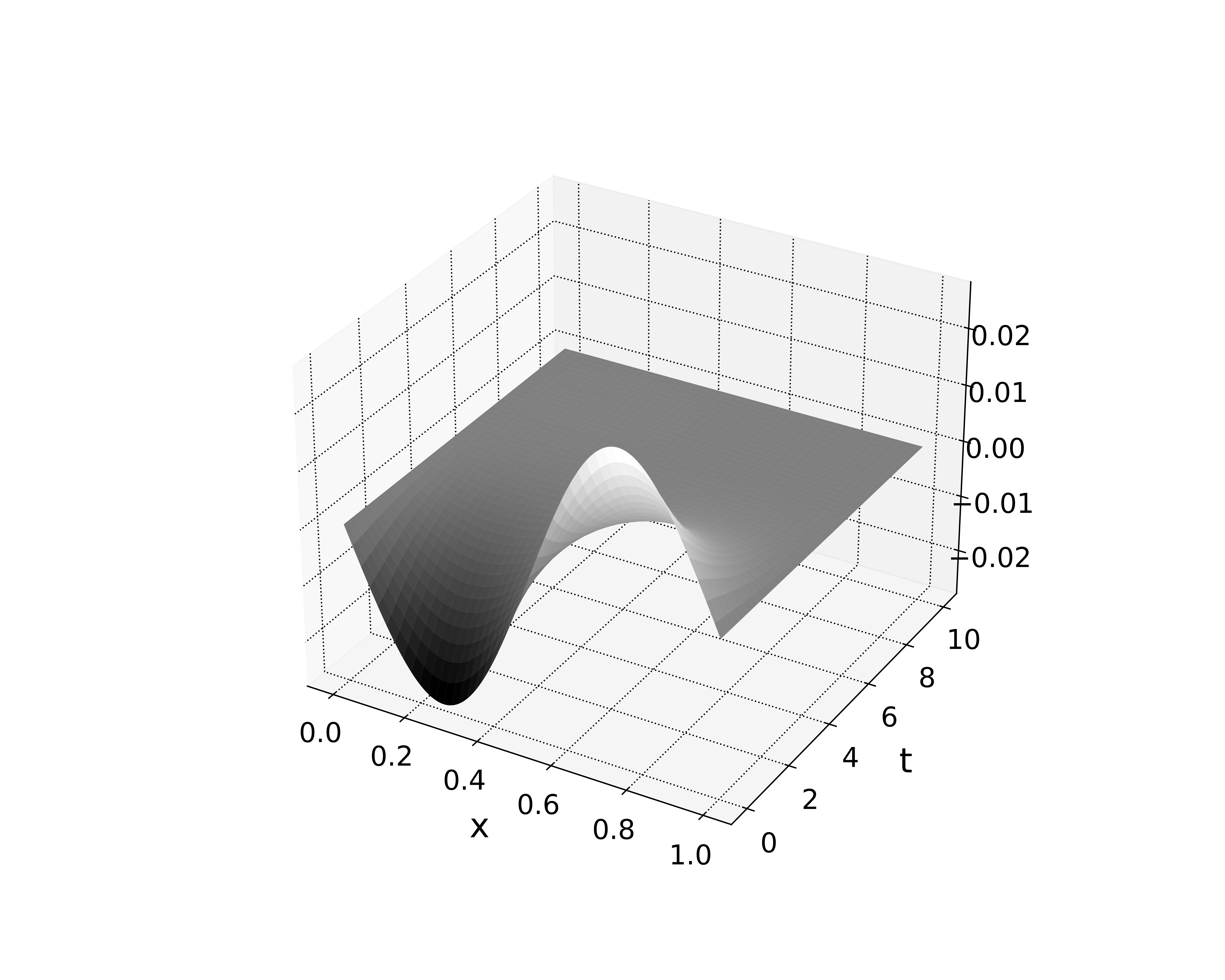}}%
	\hfill 
	\subcaptionbox{$\alpha_I$ for EGC model}{\includegraphics[width=0.33\linewidth]{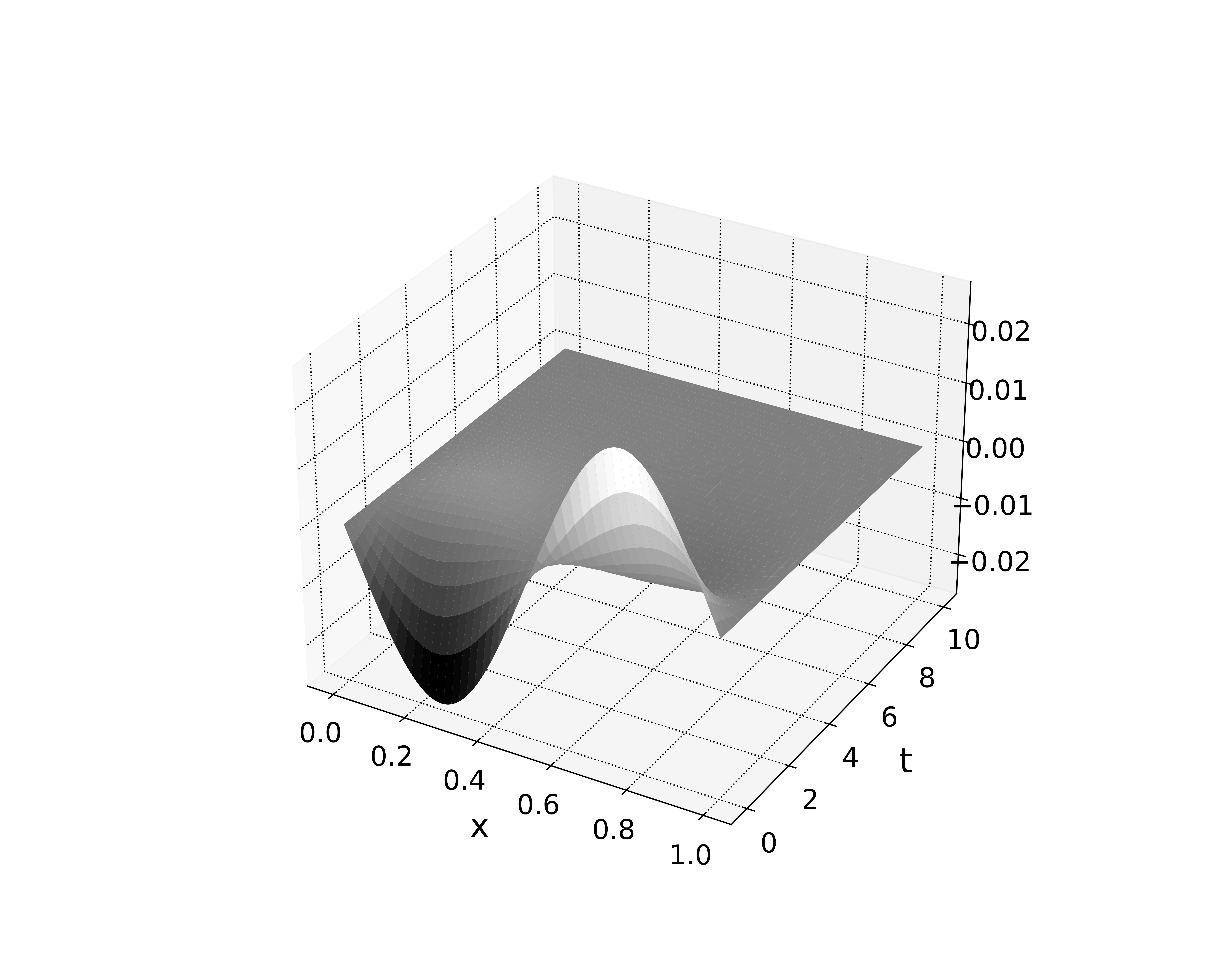}}%
	\caption{Comparison of result controls obtained from mean field models with cost functional $J_{EGC}^{2}$ and $J_{TGC}^{2}$, which considers running and current costs}
	\label{fig_alpha_only_currCost}
\end{figure}

The optimal strategies obtained for functionals $J_{EGC}^{2}$ and $J_{TGC}^{2}$ are shown in Figure \ref{fig_alpha_only_currCost}. Note that the ``view'' of strategies for the TGC and EGC models are the same (since terminal conditions are not used here), but in absolute terms, the value spread of strategy $\alpha$ of the TGC model is more than twice the sum of strategies $\alpha_S$ and $\alpha_I$ for EGC model. This has an impact on the forecast results. Table \ref{tab_1} shows the root-mean-square difference between the results obtained according to the differential SIR, SIR EGC MF and SIR TGC MF models over the forecast period $T=10$ days. Note that here and below strategy $\alpha_R$ is identically equal to zero for the SIR EGC MF models defined by functionals $J^l_{EGC},\;l=1,...,4$, therefore, in order not to clutter the description, we will not give the corresponding figures for $\alpha_R$.

\begin{table}[H]
	\caption{Standard deviation (in people) calculated for a 10-day period between the models SIR, SIR EGC, SIR TGC, determined by the functionals: $J_{EGC}^{2}, J_{TGC}^{2}$ (Current cost); $J_{EGC}^{3}, J_{TGC}^{3}$ (Terminal cost); $J_{EGC}^{4}, J_{TGC}^{4}$ (Current + Terminal costs)  }
	\label{tab_1}
	\begin{tabular}{|p{2cm}|p{0.5cm}p{0.5cm}p{0.5cm}|p{0.5cm}p{0.5cm}p{0.5cm}|p{0.5cm}p{0.5cm}p{0.5cm}|}
		\hline
		\multirow{2}{*}{models} & \multicolumn{3}{c|}{Current cost}                      & \multicolumn{3}{c|}{Terminal cost}                     & \multicolumn{3}{c|}{Current+Terminal costs}            \\ \cline{2-10} 
		& {S}  & {I}  & R  & {S}  & {I}  & R  & {S}  & {I}  & R  \\ \hline
		TGC-EGC                 & {4}  & {2}  & 3  & {4}  & {2}  & 3  & {3}  & {1}  & 2  \\ \hline
		EGC-SIR                 & {15} & {8}  & 11 & {30} & {15} & 22 & {18} & {10} & 13 \\ \hline
		TGC-SIR                 & {19} & {10} & 14 & {26} & {13} & 19 & {20} & {11} & 15 \\ \hline
	\end{tabular}
\end{table}

\begin{figure}[H]
	\centering
	\subcaptionbox{$\alpha$ for TGC model}{\includegraphics[width=0.33\linewidth]{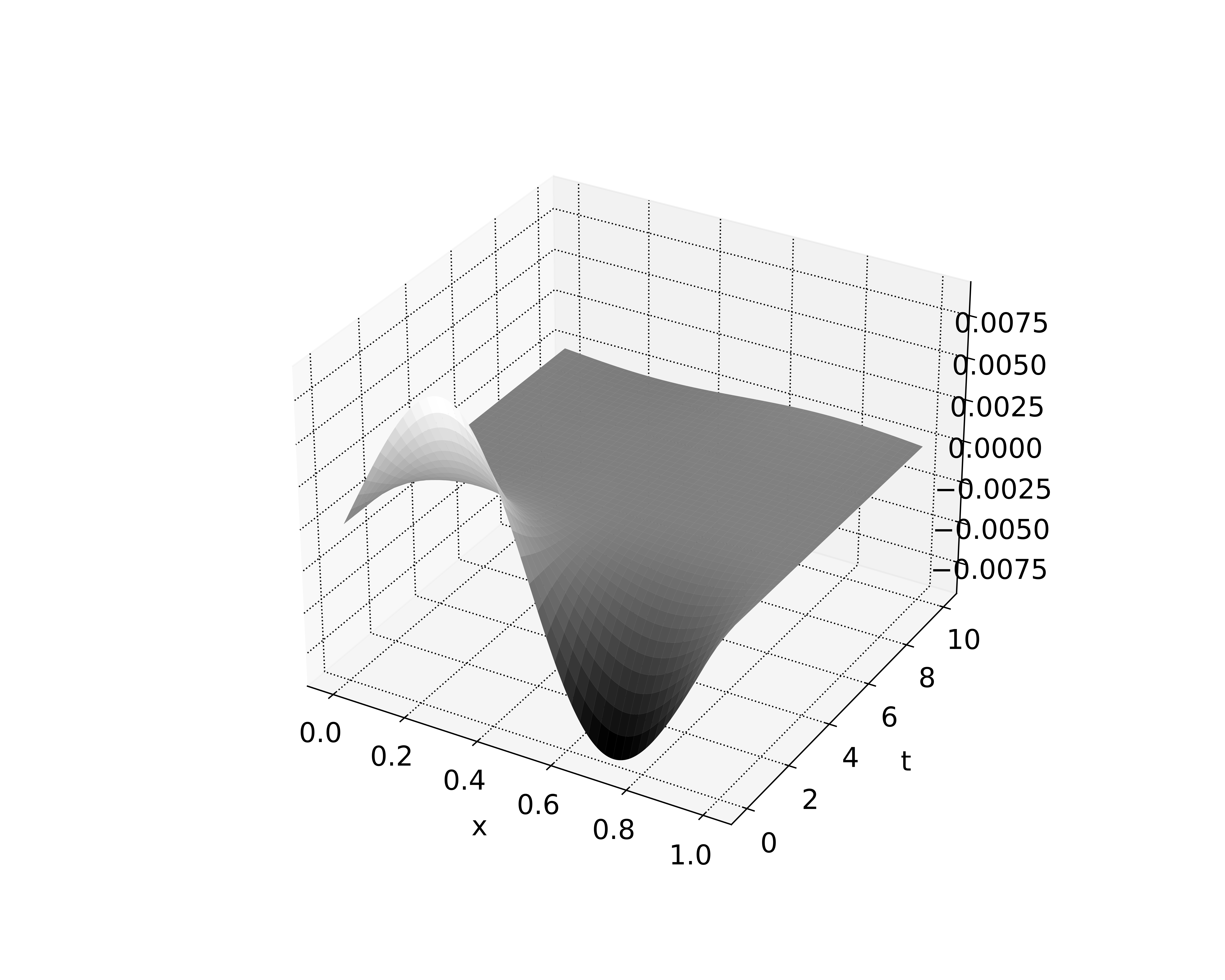}}%
	\hfill 
	\subcaptionbox{$\alpha_S$ for EGC model}{\includegraphics[width=0.33\linewidth]{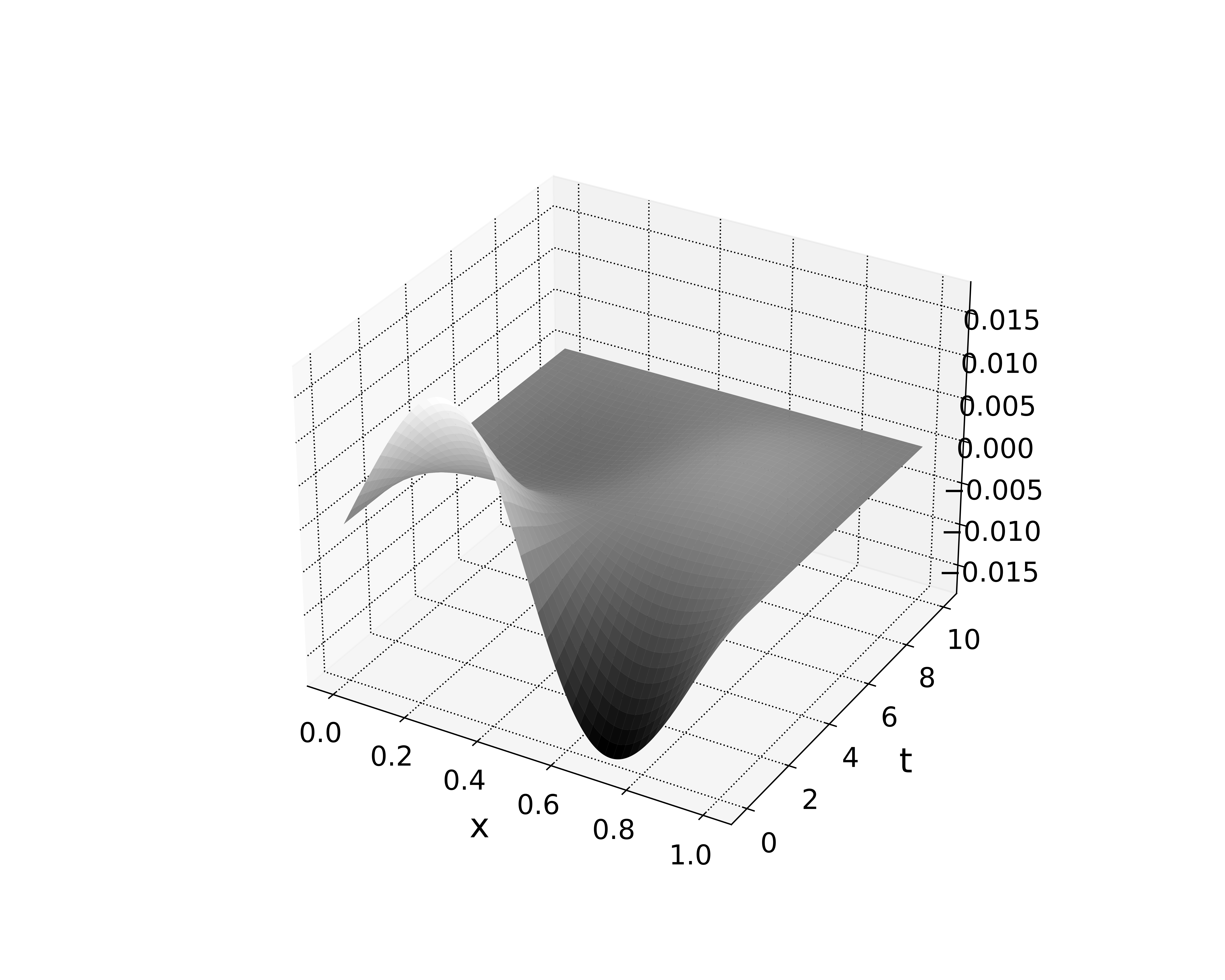}}%
	\hfill 
	\subcaptionbox{$\alpha_I$ for EGC model}{\includegraphics[width=0.33\linewidth]{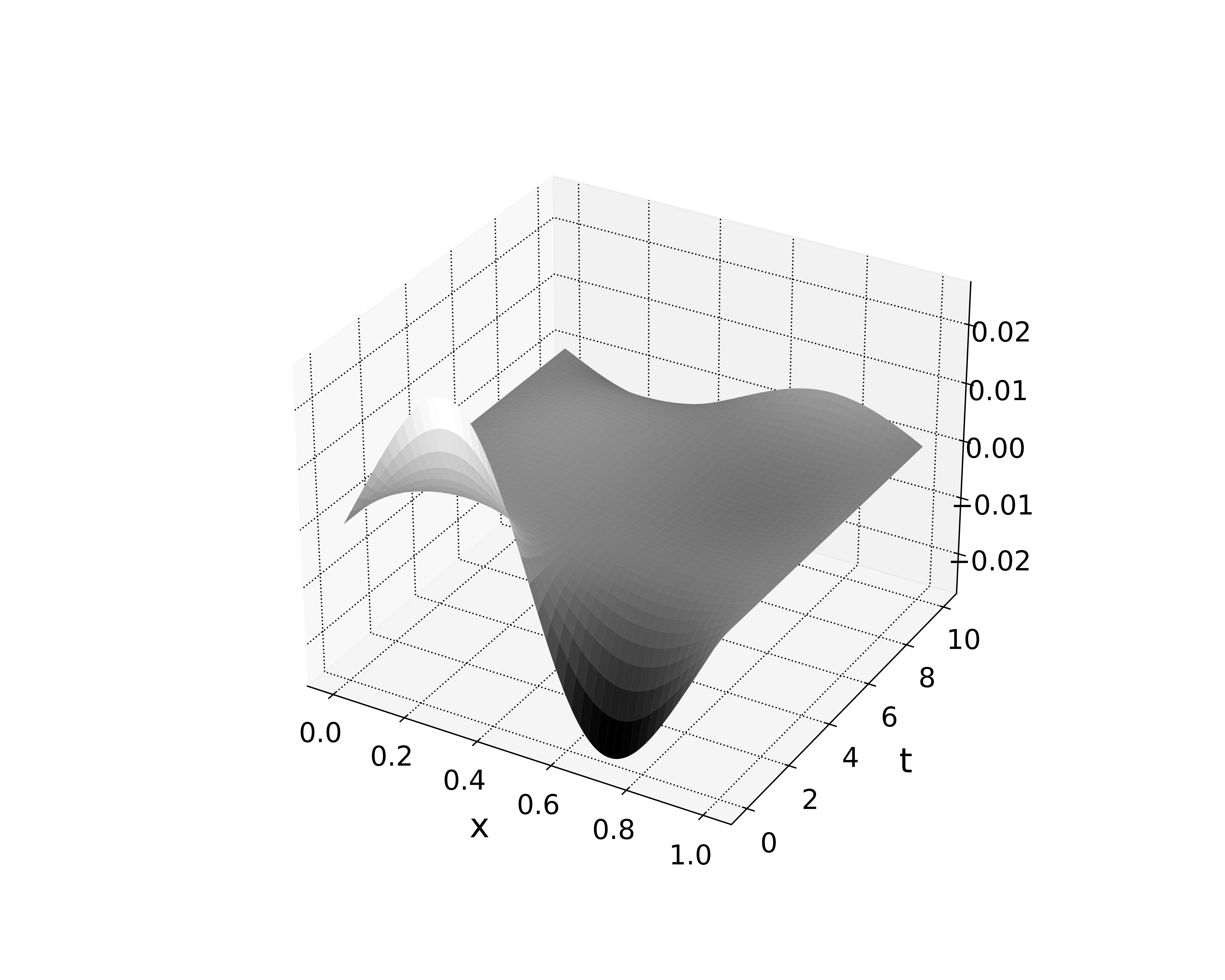}}%
	\caption{Comparison of result controls obtained from mean field models with cost functional $J_{EGC}^{3}$ and $J_{TGC}^{3}$, which considers running and terminal costs}
	\label{fig_alpha_only_TerminalCost}
\end{figure}

Now let us consider  how terminal conditions influence the obtained type of optimal control (models with function $J^3_{EGC},J^3_{TGC}$). From Figure \ref{fig_alpha_only_TerminalCost} it is clear that the imposed terminal conditions on only one epidemiological group significantly influence the strategy of this group in the SIR EGC MF model. At the same time, for the TGC model this difference is blurred due to the generality of control for all epidemiological groups. The contribution of terminal conditions makes the most difference in the modeling process (see table \ref{tab_1}).

\begin{figure}[H]
	\centering
	\subcaptionbox{$\alpha$ for TGC model}{\includegraphics[width=0.33\linewidth]{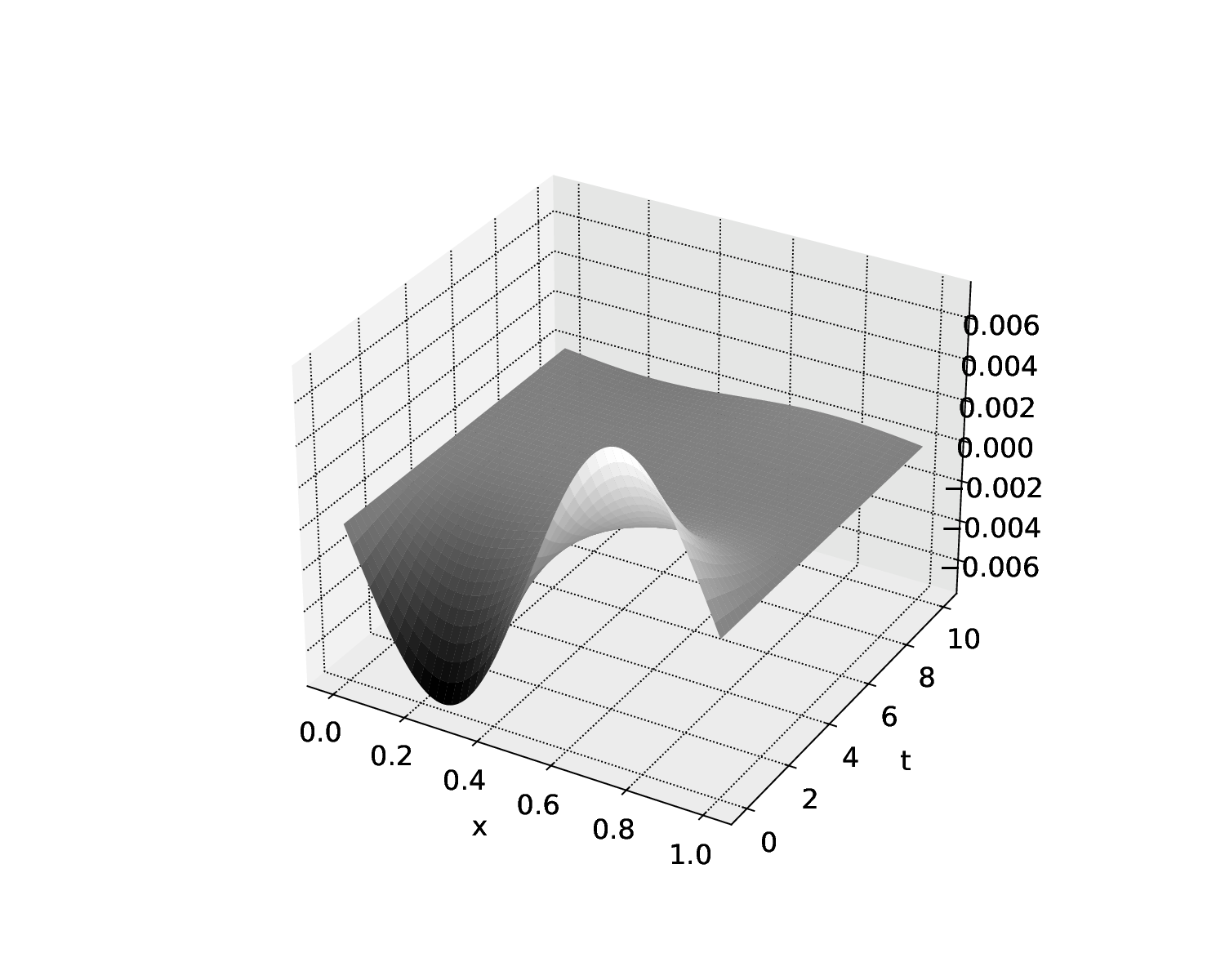}}%
	\hfill 
	\subcaptionbox{$\alpha_S$ for EGC model}{\includegraphics[width=0.33\linewidth]{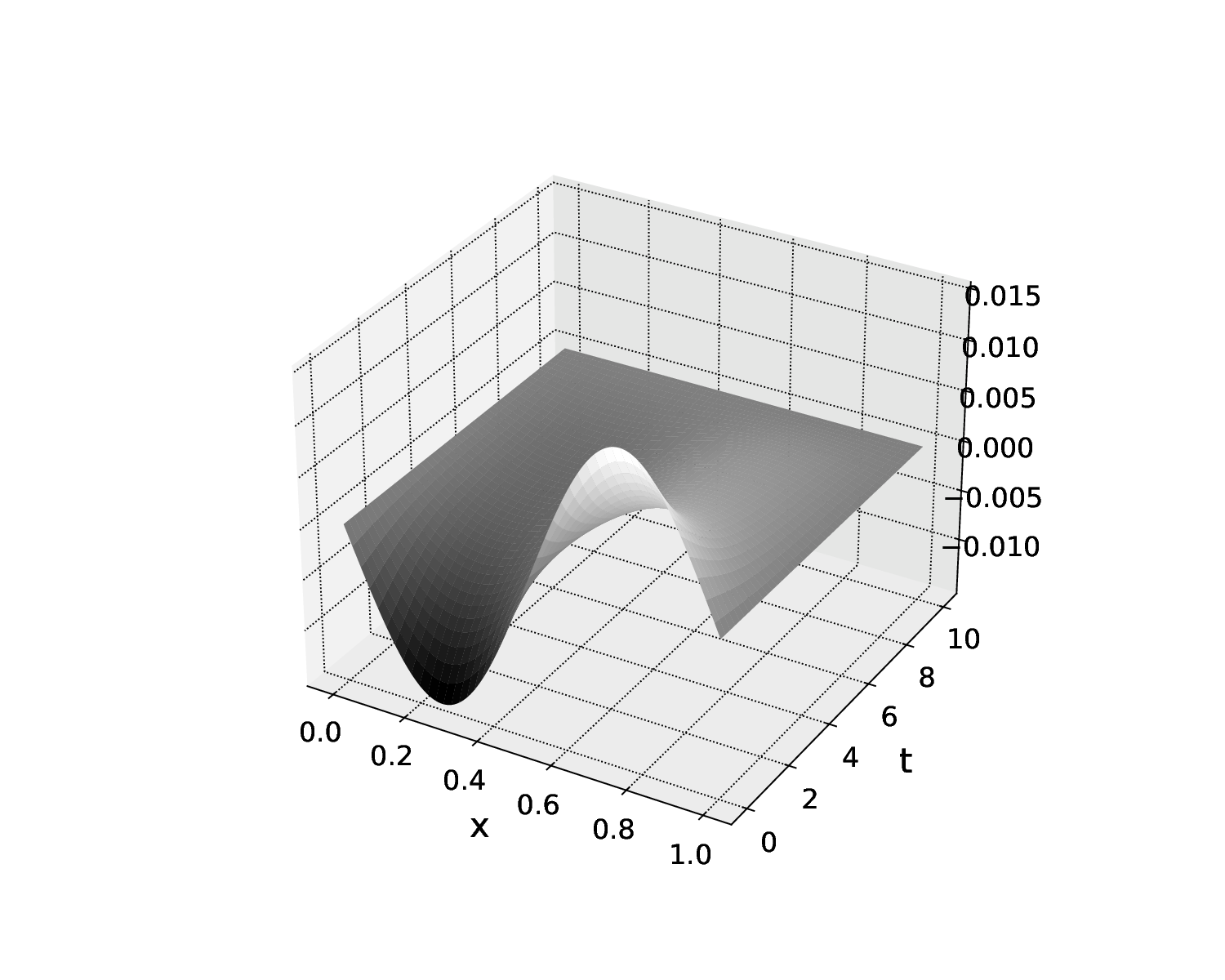}}%
	\hfill 
	\subcaptionbox{$\alpha_I$ for EGC model}{\includegraphics[width=0.33\linewidth]{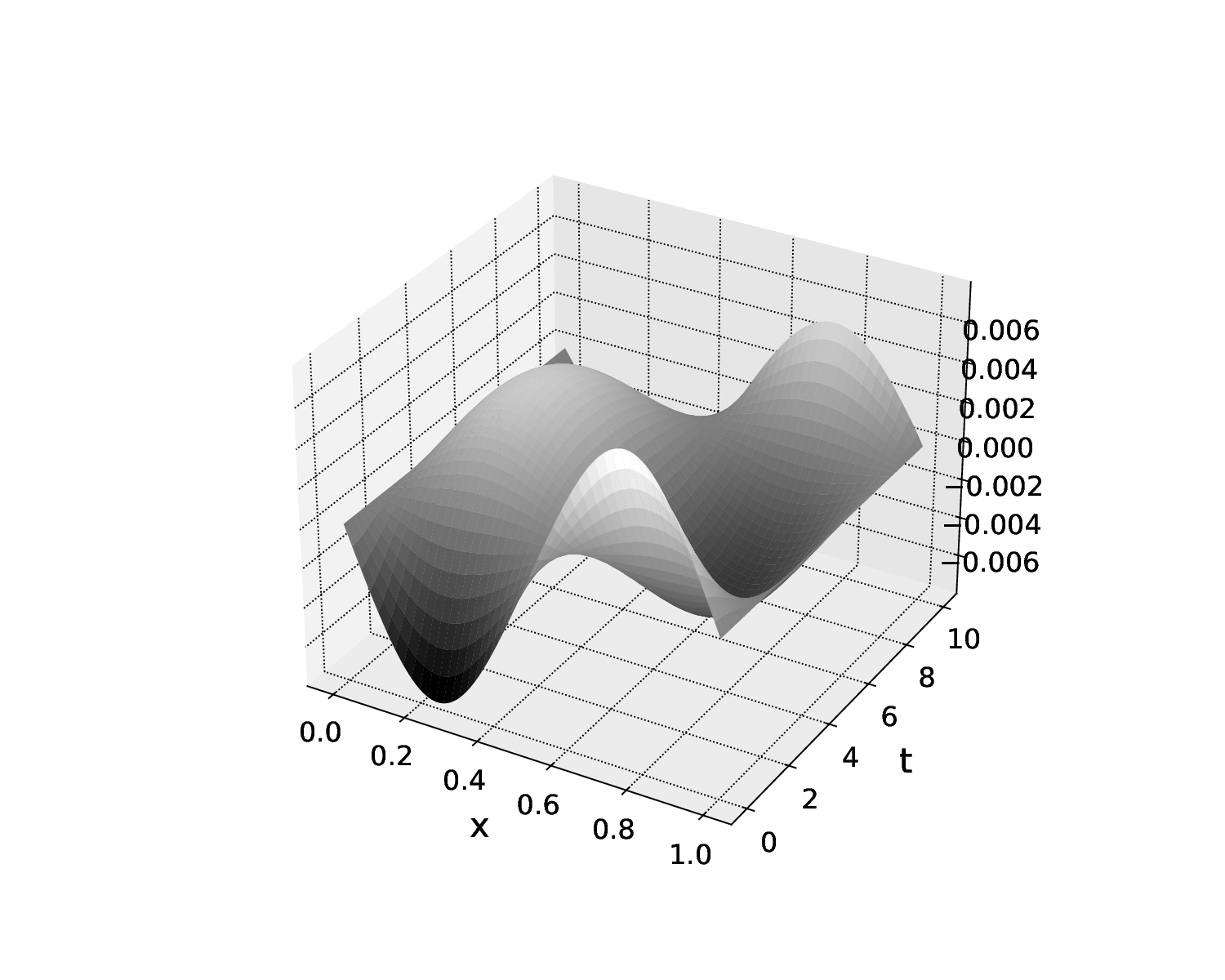}}%
	\caption{Comparison of result controls obtained from mean field models with cost functional $J_{EGC}^{4}$ and $J_{TGC}^{4}$, which considers running, current and terminal costs}
	\label{fig_alpha_Curr_TerminalCost}
\end{figure}

And finally, Figure \ref{fig_alpha_Curr_TerminalCost} shows how great the smoothing of optimal control is (models with functionals $J^4_{EGC}$ and $J^4_{TGC}$). 

An example of simulation results is presented in Figure \ref{fig_m_for_only_TerminalCost}.

According to the numerical analysis, even minor differences in models, such as an isolation strategy applied by the population as a whole or separately, can significantly influence the final result. However,  the difference between the mean-field models obtained in the model example is not large, which can be explained by several reasons. Firstly, the problem has been modeled, so the considered population is not too large. In real problems, where the population is millions of  people, a relative error of 0.1\% can amount to several thousand people. Secondly, the modeling period of 10 days is not too long and the differences between the models do not have time to affect the behavior of the system. Thirdly, the system’s temporal dynamics is depends more on epidemiological parameters  rather than on the  chosen  isolation strategy.

\section{An example of using presented formulations to model the spread of COVID-19} \label{sec_example_real}

\subsection{Basic differential model}

\begin{figure}[h]
	\centering
	\includegraphics[width=0.7\linewidth]{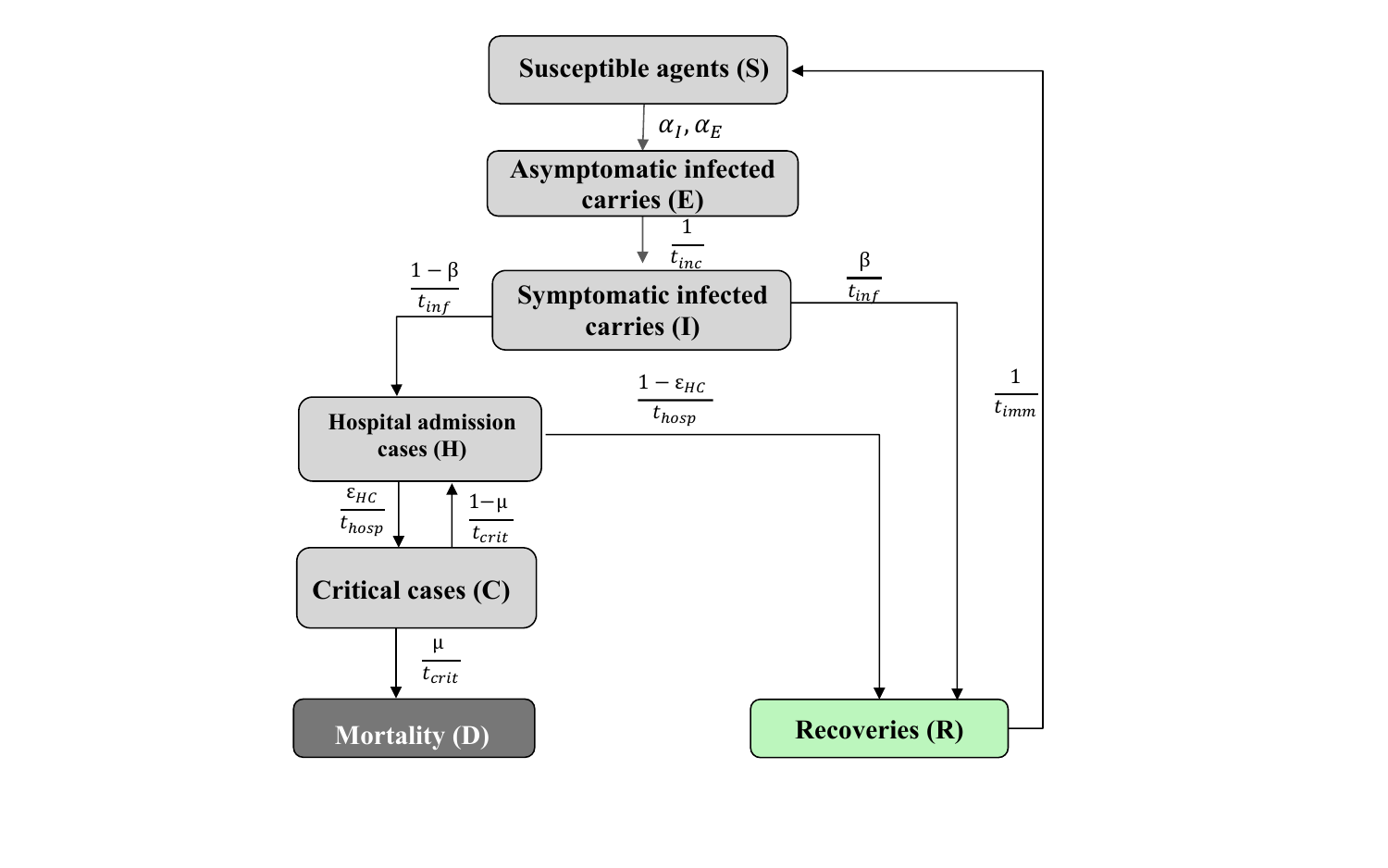}
	\caption{SEIR-HCD flow diagram}
	\label{fig_SEIRHCD_flow_diagramm}
\end{figure}

Finally, we will compare the described models as applied to the simulation of a real epidemiological situation. As a basic differential model, instead of SIR, we will use the SEIR-HCD model, described in work \cite{Krivorotko_SEIRHCD_2022} and represented by a system of differential equations
\begin{equation}
	\label{eq_SEIRHCD_differential}
	\left\{
	\begin{aligned}
		& {d m_S}\big/{dt} = -{\left(5-a\right)} \left({\alpha_I(t)m_S(t)m_I(t)}+{\alpha_E(t)m_S(t)m_E(t)}\right)\big/{5} + {\omega_{imm}}m_R(t),\\
		& {d m_E}\big/{dt} = {\left(5-a\right)} \left({\alpha_I(t)m_S(t)m_I(t)}+{\alpha_E(t)m_S(t)m_E(t)}\right)\big/{5} - {\omega_{inc}}m_E(t),\\
		& {d m_I}\big/{dt} = {\omega_{inc}}m_E(t) - {\omega_{inf}}m_I(t),\\
		& {d m_R}\big/{dt} = {\beta}{\omega_{inf}}m_I(t) + {\left(1-\varepsilon_{HC}\right)}{\omega_{hosp}}m_H(t) - {\omega_{imm}}m_R(t),\\
		& {d m_H}\big/{dt} = {\left(1-\beta\right)}{\omega_{inf}}m_I(t) + {\left(1-\mu\right)}{\omega_{crit}}m_C(t) - {\omega_{hosp}}m_H(t),\\
		& {d m_C}\big/{dt} = {\varepsilon_{HC}}{\omega_{hosp}}m_H(t) - {\omega_{crit}}m_C(t),\\
		& {d m_D}\big/{dt} = {\mu}{\omega_{crit}}m_C(t)\\
	\end{aligned}
	\right.
\end{equation}
with initial values for each in the set \{$m_i(0)=m_{i0}\;i \in \{S,E,I,R,H,C,D\}$. Here the population is divided into seven clusters: $S(t)$  is the not immune part of population, $E(t)$ denoted number of individuals who asymptomatically infected, $I(t)$ -- symptomatically infected, $R(t)$ denotes recovered or immune part of population, $H(t)$ -- people who are hospitalized, $C(t)$ denotes critically-ill individuals on mechanical ventilation, $D(t)$ -- dead due to COVID-19. The groups are paired by probabilities (model coefficients) and transfer from one group to another is performed in accordance Fig. \ref{fig_SEIRHCD_flow_diagramm}. 

The description of model parameters is presented in Table \ref{tab_1}. In contrast to the flow diagram \ref{fig_SEIRHCD_flow_diagramm} instead of the number of people in each group of population, we introduce the fraction of each epidemiological group at the time moment $t$ and introduce the frequencies $\omega_{imm},\omega_{inc}, \omega_{inf}, \omega_{hosp}, \omega_{crit}$, where $\omega_{\circ} = 1 \big/ {t_{\circ}}$  with corresponding values instead $\circ$.

\begin{table}[h]
	\caption{\label{tab_2}Description of SEIR-HCD model parameters}
	\centering
	\begin{tabular}{|p{30pt}|p{300pt}|}
		\hline
		\textbf{Symbol} & \textbf{Description} \\
		\hline
		$a$ & Yandex self-isolation index \\
		\hline
		$\alpha_E$ & Infection parameter for asymptomatic and susceptible groups \\
		\hline
		$\alpha_I$ & Infection parameter for infected and susceptible groups \\
		\hline
		$\beta$ & Portion of infected cases with no complications \\
		\hline
		$\varepsilon_{HC}$ & Portion of hospitalized cases on mechanical ventilation \\
		\hline
		$\mu$ & Mortality rate due to COVID-19 \\
		\hline
		$t_{inc}$ & The number of days since a contact before an agent becomes contagious \\
		\hline
		$t_{inf}$ & The duration of infectious with symptoms \\
		\hline
		$t_{hosp}$ & The number of days for a severe case to become a critical one \\
		\hline
		$t_{crit}$ & The duration of critical condition \\
		\hline
		$t_{imm}$ & The duration of immunity to COVID-19 \\
		\hline
	\end{tabular}
\end{table}

We use the SEIR-HCD model for several reasons. Firstly, the SIR model, as a rule, is not used to solve real modeling problems due to a rather superficial description of the epidemiological process. In addition, the SEIR-HCD model provides a precise portrayal of the population, considering the potential loss of immunity over time, providing an advantage in itself.  A thorough analysis of the model using \cite{Krivorotko_SEIRHCD_2022} was conducted, including estimates of the intervals where the various parameters vary, algorithms for determining these parameters relative to statistical data, and a sensitivity analysis using the Sobol method. The SEIR-HCD model was effectively utilized to explain the epidemiological scenarios in Novosibirsk \cite{Krivorotko_2020,Krivorotko_SEIRHCD_2020} and Moscow \cite{Krivorotko_2020}.

\subsection{Mean field models}

To save space, we will not describe in detail the view of the SEIR-HCD EGC and TGC MF models. Their construction does not differ from that presented in Sections \ref{sec_formulation} in continuous form and \ref{sec_finite_diff} for discrete ones. Here we will only present the functionals  in accordance with which the modeling was carried out. Thus, as EGC MF optimization probles with SEIR-HCD clusterization we will consider the problem of the corresponding functional minimization:
\begin{equation}
	\label{eq_EEGC_SEIRHCD_func}
	\begin{aligned}
		J_{EGC}^{SEIRHCD}=\int_{0}^{T}&\int_{0}^{1}
		(\underset{i\in\{S,E,I,R,H,C,D\}}{\sum}\frac{\alpha_i^2 m_i}{2} +  d_1 (m_E^2 + m_I^2 + \left(1-m_R\right)^2 \\
		&  + \left(1-m_D\right)^2 ))\text{d}x\text{d}t + d_2\int_0^1 m_I^2(T,x) \text{d}x.
	\end{aligned}
\end{equation}
The functional $J_{TGC}^{SEIRHCD}$ is obtained from \eqref{eq_EEGC_SEIRHCD_func} bu replacing $\alpha_i$ on $\alpha$.

\subsection{Computational experiment}

\begin{figure}[H]
	\centering
	\subcaptionbox{S group}{\includegraphics[width=0.33\textwidth]{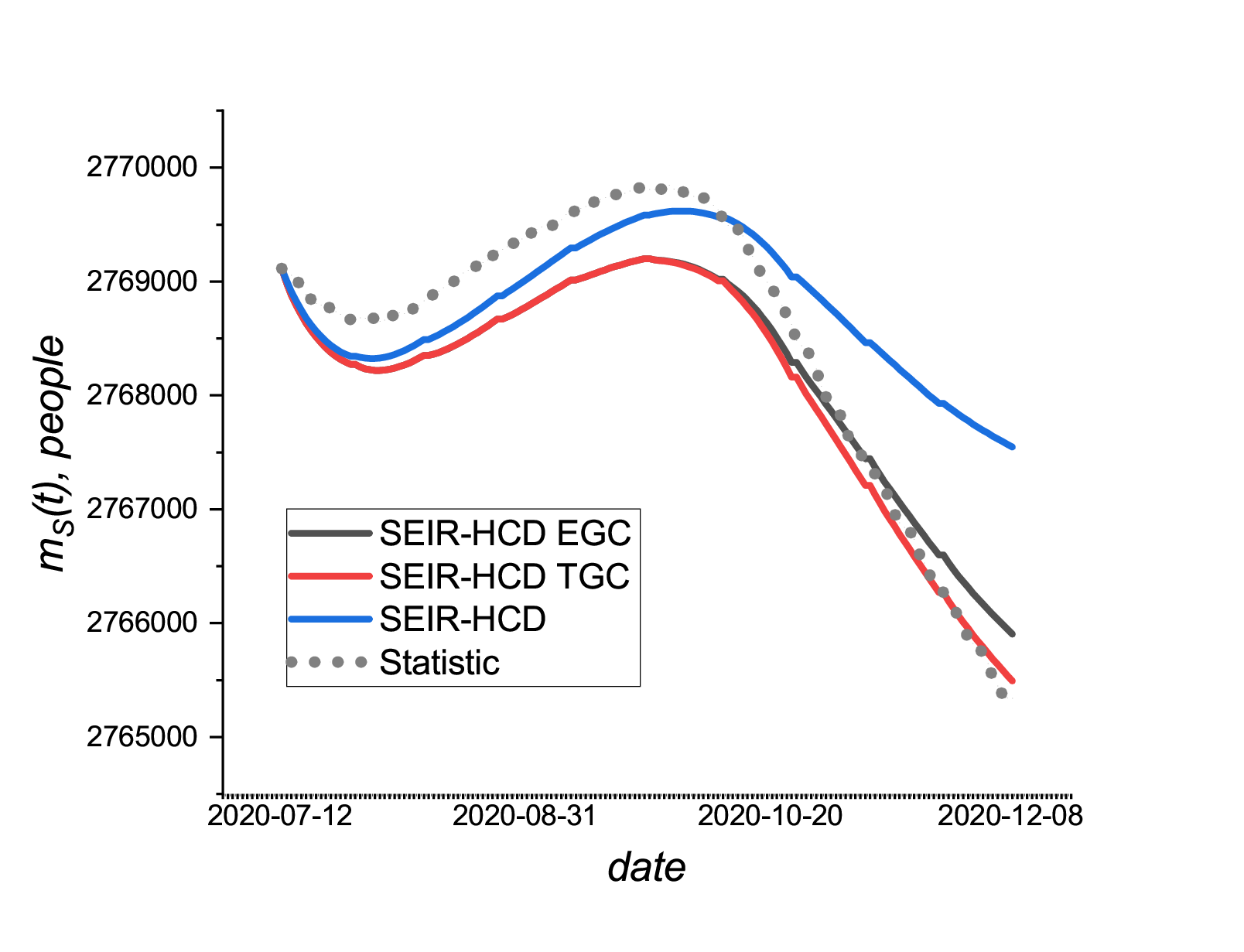}}%
	\hfill 
	\subcaptionbox{E group}{\includegraphics[width=0.33\textwidth]{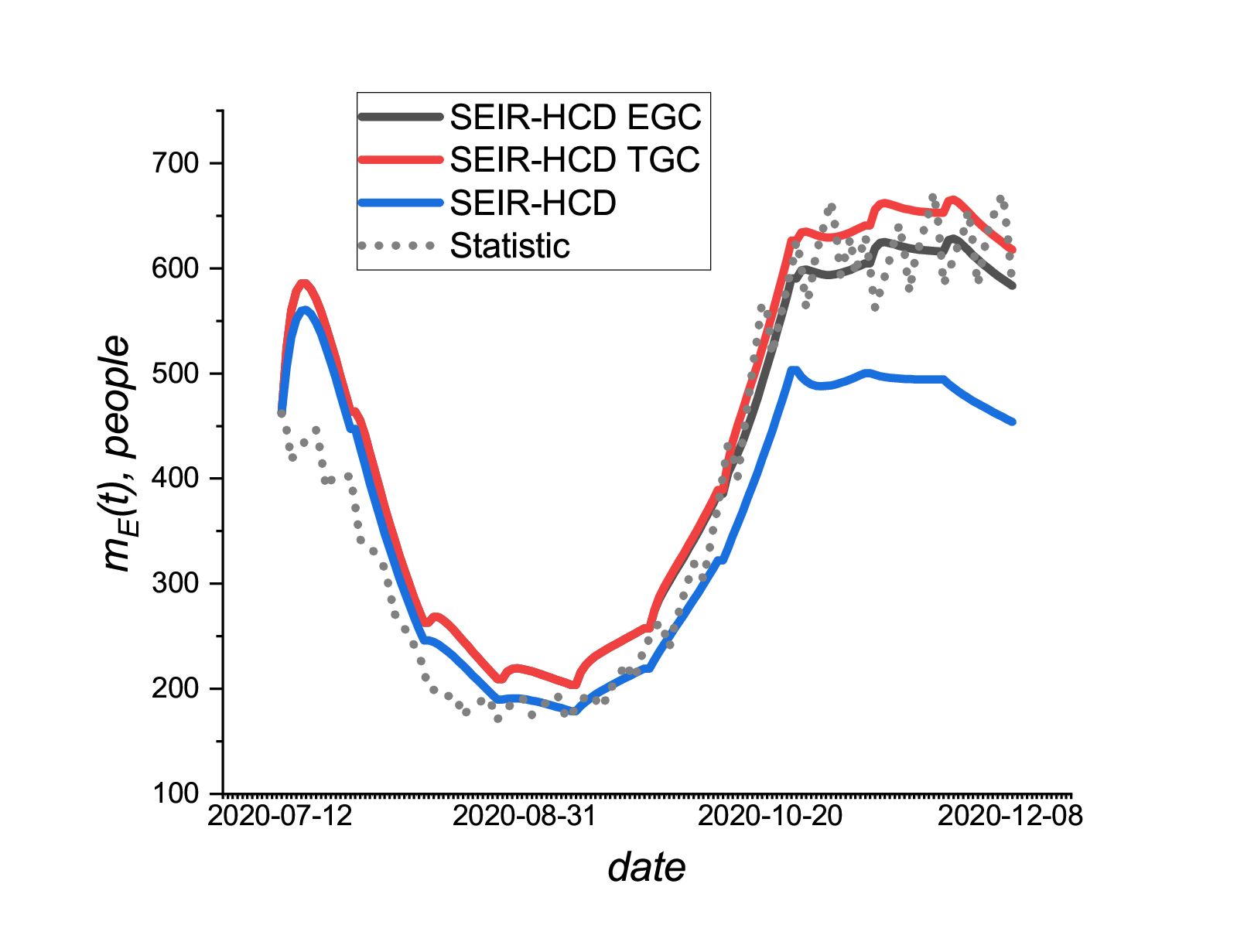}}%
	\hfill 
	\subcaptionbox{I group}{\includegraphics[width=0.33\textwidth]{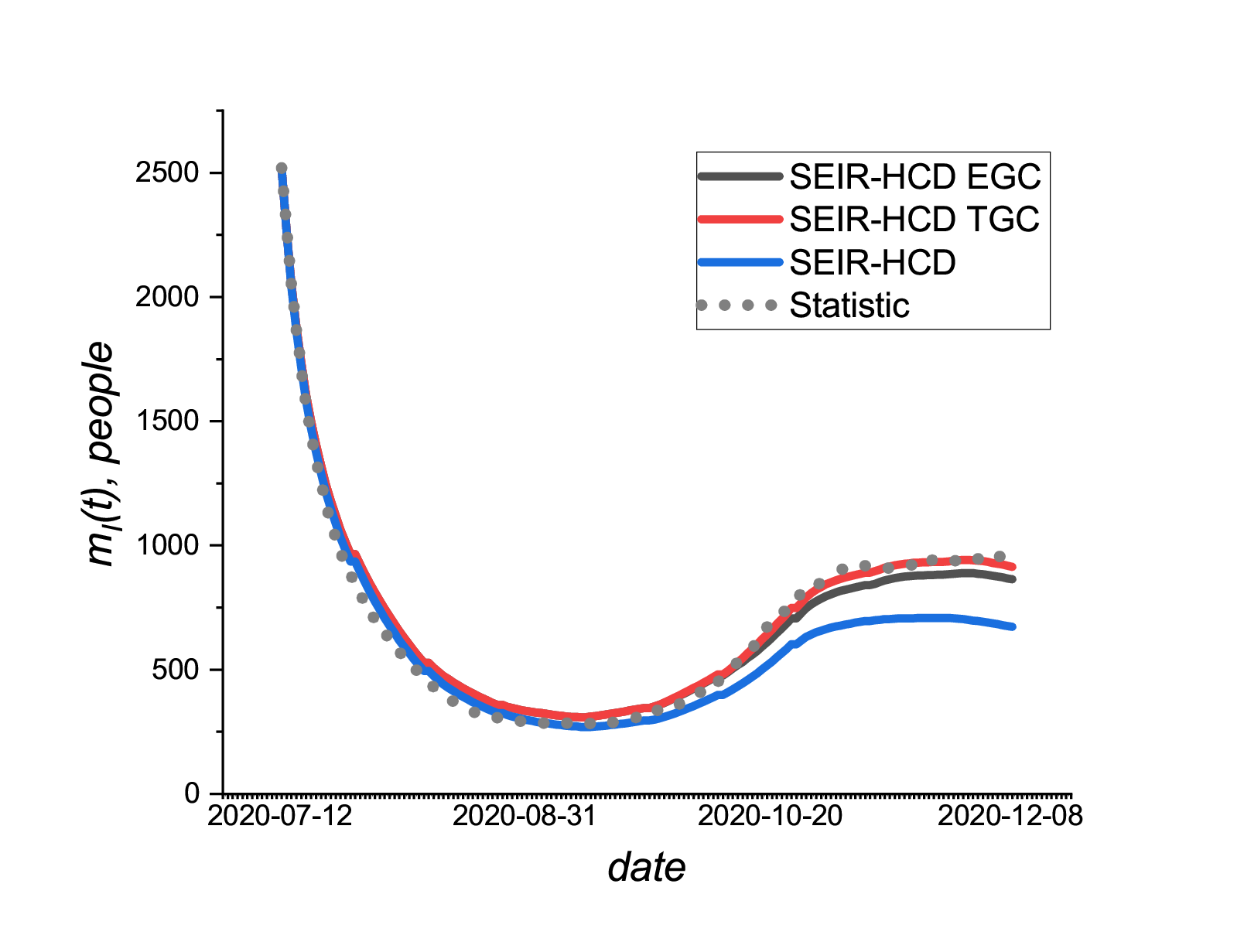}}%
	\hfill
	\subcaptionbox{R group}{\includegraphics[width=0.33\textwidth]{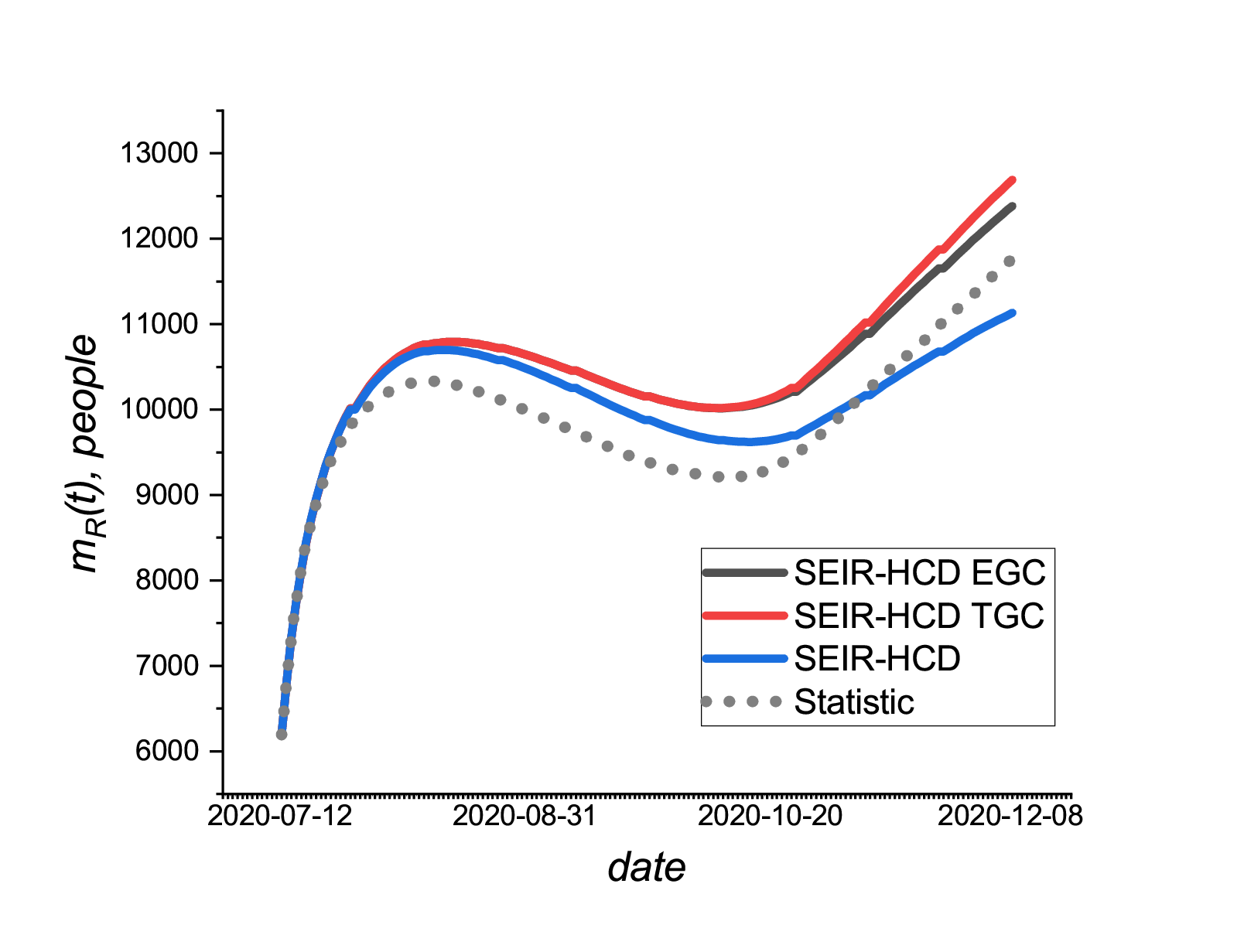}}%
	\hfill 
	\subcaptionbox{H group}{\includegraphics[width=0.33\textwidth]{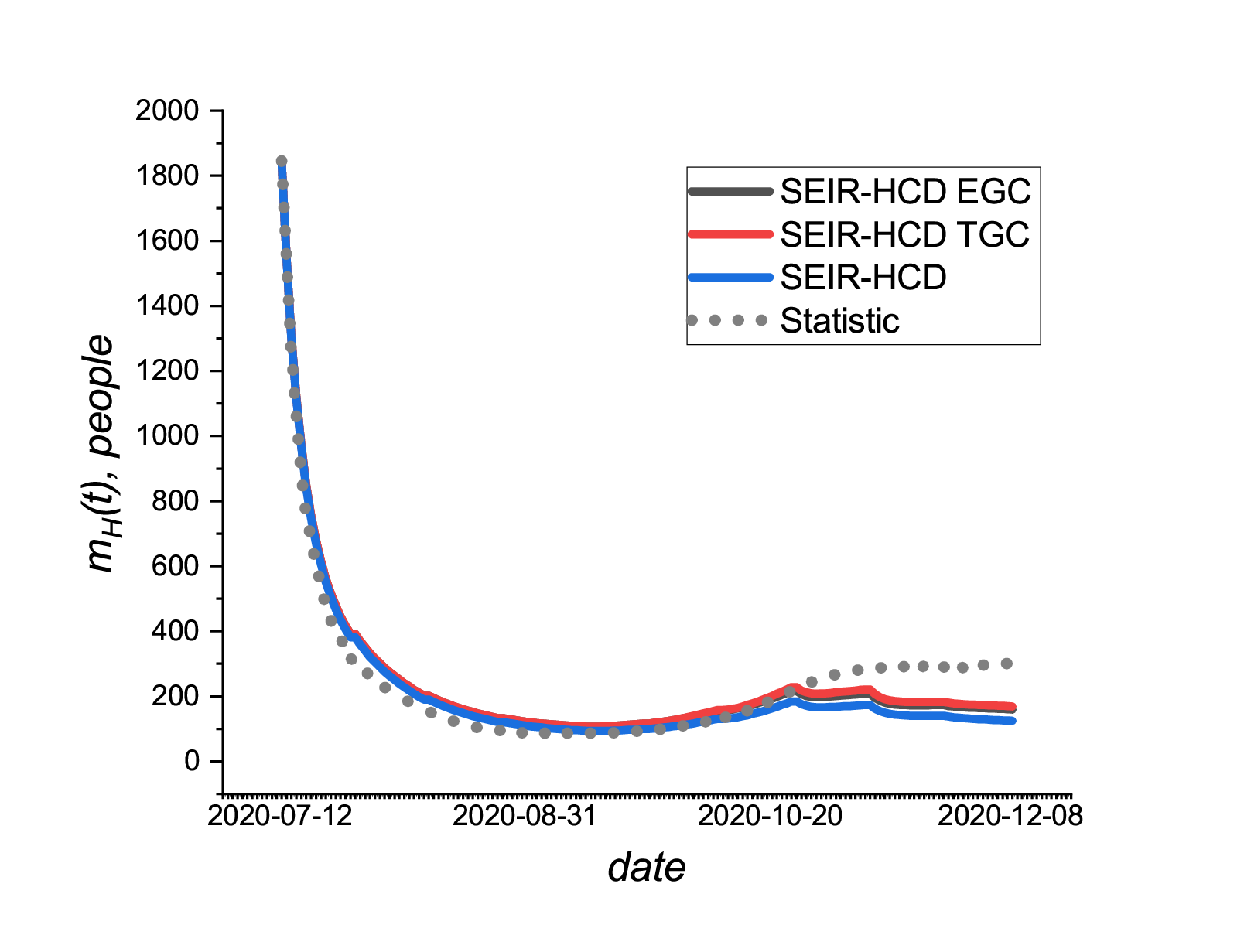}}%
	\hfill 
	\subcaptionbox{C group}{\includegraphics[width=0.33\textwidth]{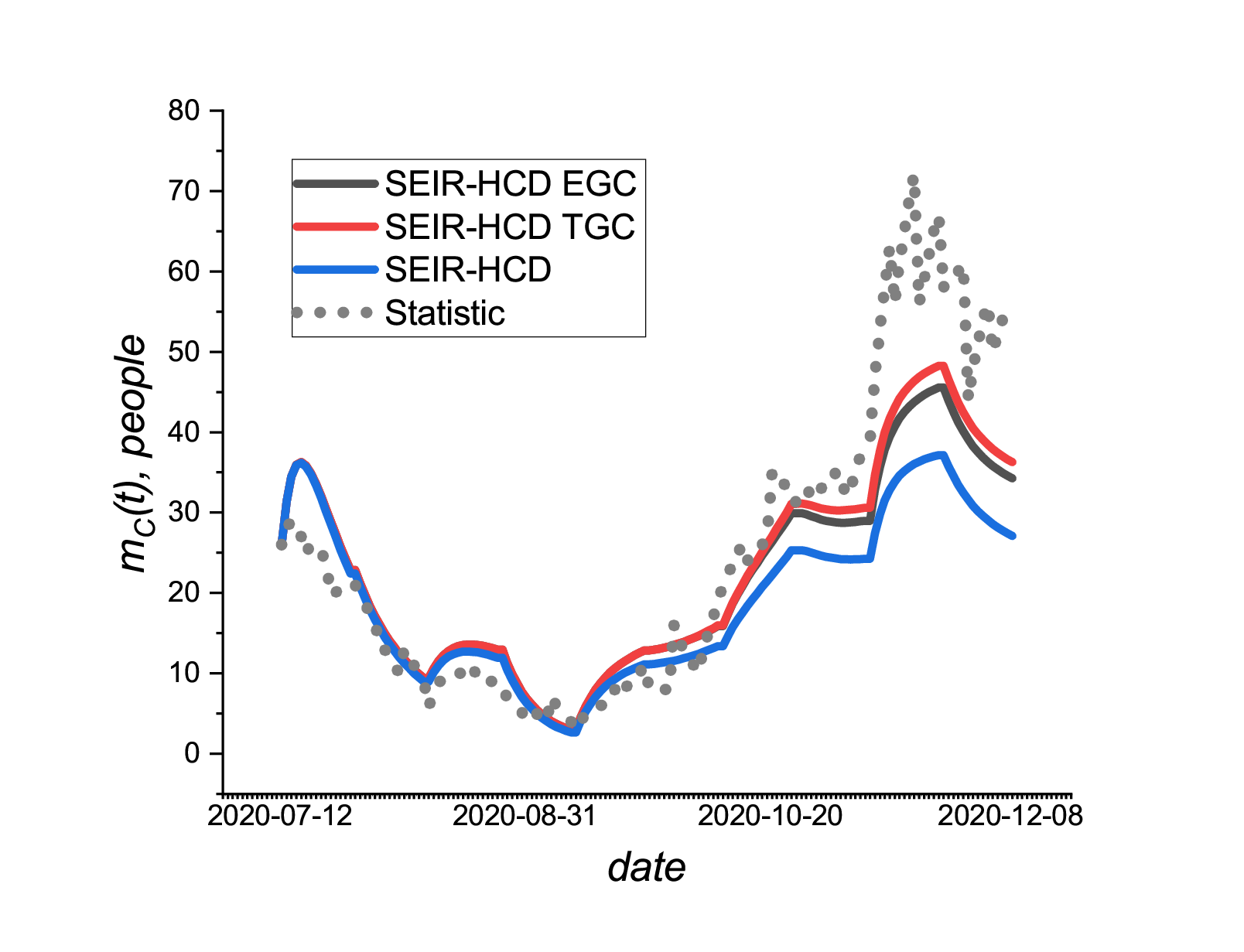}}%
	\caption{Comparison of simulation results of differential SEIR-HCD, SEIR-HCD EGC MF, SEIR-HCD TGC MF with real data in Novosibirsk in 2020 with simulation window ($w$) equaled to 15 days }
	\label{fig_SEIRHCD_modelling_result}
\end{figure}

We will compare the modeling results with statistics on the incidence of Covid-19 in Novosibirsk for 150 days from 2020-07-12. The corresponding statistics for this city and some others are presented on the website \url{https://covid19-modeling.ru/data}. The values of the parameters of the SEIR HSD model, described in the table \ref{tab_2}, for each day of modeling are presented in the file \url{https://disk.yandex.ru/i/O9jtRV-xEV3tMA}. These parameters were obtained by a group led by O.I. Krivorotko according to the methods described in the work \cite{Krivorotko_SEIRHCD_2022}. For the computational experiment, we will also introduce the concept of ``simulation window'' ($w$) - the period over which the parameters were averaged. Let's explain with an example. If $w$ is equal to 7, then for the model we consider that the simulation is carried out for the time period $T=7$, and the epidemiological parameters described in the table \ref{tab_2} and used for the current simulation are averaged over the corresponding time period (7 days). Between modeling periods, gluing is performed. For other parameters not described in the attached file, put
\[\sigma_i=0.02, x_i^c = 0.5,\; \sigma_i^c = 0.2,\; d_1 = 10^{-5},\; d_2 = 10.\]
The population in Novosibirsk for the specified modeling period will be considered equal to 2780288, and the parameter $w$ is chosen from the set $w=\{50,30,15,10,5,3\}$ days. The simulation results are presented in Figure \ref{fig_SEIRHCD_modelling_result} for a simulation window of 15 days. The root mean square error obtained for all models with different values of $w$ is presented in the table \ref{tab_3}.

\begin{table}[h]
	\caption{\label{tab_3} Root mean square difference of simulation results with real data (in people) for different simulation windows}
	\centering
	\begin{tabular}{|llllllll|}
		\hline
		\multicolumn{1}{|l|}{model}    & \multicolumn{1}{l|}{S}    & \multicolumn{1}{l|}{E}   & \multicolumn{1}{l|}{I}   & \multicolumn{1}{l|}{R}    & \multicolumn{1}{l|}{H}  & \multicolumn{1}{l|}{C}  & D   \\ \hline
		\multicolumn{8}{|c|}{w=50 days}                                                                                                                                                                        \\ \hline
		\multicolumn{1}{|l|}{SEIR-HCD} & \multicolumn{1}{l|}{862}  & \multicolumn{1}{l|}{113} & \multicolumn{1}{l|}{150} & \multicolumn{1}{l|}{875}  & \multicolumn{1}{l|}{85} & \multicolumn{1}{l|}{14} & 391 \\ \hline
		\multicolumn{1}{|l|}{EGC}      & \multicolumn{1}{l|}{980}     & \multicolumn{1}{l|}{108}    & \multicolumn{1}{l|}{134}    & \multicolumn{1}{l|}{1112}     & \multicolumn{1}{l|}{82}   & \multicolumn{1}{l|}{13}   & 386     \\ \hline
		\multicolumn{1}{|l|}{TGC}      & \multicolumn{1}{l|}{981}  & \multicolumn{1}{l|}{107} & \multicolumn{1}{l|}{133} & \multicolumn{1}{l|}{1114} & \multicolumn{1}{l|}{82} & \multicolumn{1}{l|}{13} & 386 \\ \hline
		\multicolumn{8}{|c|}{w=30 days}                                                                                                                                                                        \\ \hline
		\multicolumn{1}{|l|}{SEIR-HCD} & \multicolumn{1}{l|}{971}  & \multicolumn{1}{l|}{105} & \multicolumn{1}{l|}{158} & \multicolumn{1}{l|}{419}  & \multicolumn{1}{l|}{81} & \multicolumn{1}{l|}{13} & 419 \\ \hline
		\multicolumn{1}{|l|}{EGC}      & \multicolumn{1}{l|}{622}     & \multicolumn{1}{l|}{76}    & \multicolumn{1}{l|}{105}    & \multicolumn{1}{l|}{479}     & \multicolumn{1}{l|}{72}   & \multicolumn{1}{l|}{11}   & 413    \\ \hline
		\multicolumn{1}{|l|}{TGC}      & \multicolumn{1}{l|}{617}  & \multicolumn{1}{l|}{75}  & \multicolumn{1}{l|}{104} & \multicolumn{1}{l|}{479}  & \multicolumn{1}{l|}{72} & \multicolumn{1}{l|}{11} & 413 \\ \hline
		\multicolumn{8}{|c|}{w=15 days}                                                                                                                                                                        \\ \hline
		\multicolumn{1}{|l|}{SEIR-HCD} & \multicolumn{1}{l|}{840}  & \multicolumn{1}{l|}{89}  & \multicolumn{1}{l|}{135} & \multicolumn{1}{l|}{372}  & \multicolumn{1}{l|}{79} & \multicolumn{1}{l|}{12} & 416 \\ \hline
		\multicolumn{1}{|l|}{EGC}      & \multicolumn{1}{l|}{473}     & \multicolumn{1}{l|}{56}    & \multicolumn{1}{l|}{61}    & \multicolumn{1}{l|}{631}     & \multicolumn{1}{l|}{67}   & \multicolumn{1}{l|}{9}   & 407    \\ \hline
		\multicolumn{1}{|l|}{TGC}      & \multicolumn{1}{l|}{464}  & \multicolumn{1}{l|}{57}  & \multicolumn{1}{l|}{48}  & \multicolumn{1}{l|}{494}  & \multicolumn{1}{l|}{63} & \multicolumn{1}{l|}{8}  & 405 \\ \hline
		\multicolumn{8}{|c|}{w=10 days}                                                                                                                                                                        \\ \hline
		\multicolumn{1}{|l|}{SEIR-HCD} & \multicolumn{1}{l|}{782}  & \multicolumn{1}{l|}{87}  & \multicolumn{1}{l|}{133} & \multicolumn{1}{l|}{402}  & \multicolumn{1}{l|}{81} & \multicolumn{1}{l|}{12} & 419 \\ \hline
		\multicolumn{1}{|l|}{EGC}      & \multicolumn{1}{l|}{630}     & \multicolumn{1}{l|}{66}    & \multicolumn{1}{l|}{71}    & \multicolumn{1}{l|}{845}     & \multicolumn{1}{l|}{70}   & \multicolumn{1}{l|}{8}   &  407   \\ \hline
		\multicolumn{1}{|l|}{TGC}      & \multicolumn{1}{l|}{831}  & \multicolumn{1}{l|}{96}  & \multicolumn{1}{l|}{93}  & \multicolumn{1}{l|}{1038} & \multicolumn{1}{l|}{61} & \multicolumn{1}{l|}{6}  & 402 \\ \hline
		\multicolumn{8}{|c|}{w=5 days}                                                                                                                                                                         \\ \hline
		\multicolumn{1}{|l|}{SEIR-HCD} & \multicolumn{1}{l|}{678}  & \multicolumn{1}{l|}{79}  & \multicolumn{1}{l|}{125} & \multicolumn{1}{l|}{615}  & \multicolumn{1}{l|}{88} & \multicolumn{1}{l|}{11} & 422 \\ \hline
		\multicolumn{1}{|l|}{EGC}      & \multicolumn{1}{l|}{2580}     & \multicolumn{1}{l|}{332}    & \multicolumn{1}{l|}{397}    & \multicolumn{1}{l|}{2197}     & \multicolumn{1}{l|}{91}   & \multicolumn{1}{l|}{9}   & 388    \\ \hline
		\multicolumn{1}{|l|}{TGC}      & \multicolumn{1}{l|}{2515} & \multicolumn{1}{l|}{316} & \multicolumn{1}{l|}{379} & \multicolumn{1}{l|}{2167} & \multicolumn{1}{l|}{91} & \multicolumn{1}{l|}{9}  & 388 \\ \hline
	\end{tabular}
\end{table}

The results of comparing models using a real example allow to conclude that the choice of the most appropriate model in the general case is determined by the modeling window. Since the epidemiological parameters are determined from the solution of the inverse coefficient problem for the differential SEIR-HCD model, then when the length of the modeling window is reduced (i.e., a more accurate determination of the parameters), the differential model gives a more accurate approximation to the real data. The sensitivity analysis in section \ref{sec_sensativity} provides indirect evidence for this conclusion, as all the models discussed above are still the most sensitive to epidemiological parameters, but the mean field models have a greater number of these parameters. Note, however, when the accuracy of coefficient identification is not so high (with modeling windows of medium length), mean field models can smooth out the error due to the control variable, which can significantly improve the prognoses. In real problems, forecasters usually don't know the exact parameters, but can only suggest their values based on already known statistics for previous time periods.

It should also be noted that for simplicity and independence of calculations, the same functional was used here at each modeling interval (regardless of the window length), which, in general, is an incorrect approach: the functional should reflect the current epidemiological situation in the region, which is quite variable. The correct selection of functionality, in turn, allows one to obtain a more significant advantage over differential models, especially when there is no confidence in the correct identification of epidemiological parameters.

\section{Conclusion and discussion}

This paper proposes a comparison of several approaches to epidemic modeling. The first approach is based on well-known epidemiological models of the SIR type, presented in the form of ordinary differential equations. The second approach is based on the mean field model originally proposed in \cite{Lee_21}. The model we are considering (discrete EGC MF) is its finite-difference analogue and was previously studied in works \cite{Petr_SIRC_22,Petrakova_IEEE2,Petr_Sens_23} for various applications. The third approach (TGC MF) is the mean field model, modified from the EGC MF point of view on the assumption that the isolation strategy is common to the entire population. The comparison was made from several points of view: analytical analysis, sensitivity analysis regarding model parameters, numerical comparison using synthetic and real examples. Let us briefly summarize the results obtained.

First, the use of the mean field approach imposes significant restrictions on the continuous formulation to ensure the existence and uniqueness of a solution to such a problem for a relatively simple differential one. The approach described in this work is to consider not a continuous formulation, but its finite-difference analogue, for which the conditions for the existence and uniqueness of a solution are not so restrictive. We can never be sure that the solutions to the continuous and discrete problems coincide.

Second, sensitivity analysis shows that mean-field models are most sensitive to identifying epidemiological parameters. Which, it would seem, does not distinguish them from their parent differential models. But from the assessment of general sensitivity indices, it follows that for stochastic parameters and initial distributions of the population over state space, it is not their individual values that are important, but their combined use together with other parameters.

Numerical analysis of the predictive capabilities of the models showed that the discrepancies in modeling results between the basic differential model and the mean field model based on it can be quite significant. Moreover, the result of prognoses using mean field models depends not only on epidemiological parameters, but also on the choosing of a functional that describes the epidemiological situation in the region. Difference between the EGC and TGC models is mainly determined by their response to terminal conditions, and if  their absolute values are small, as well the independence of the cost of implementing the strategy relative to different epidemiological groups, then this difference is not too large. However, difference will be significant if we take control-dependent values of epidemiological parameters, for example, $\beta=\beta(\alpha)$ for TGC MF model or $\beta_i=\beta(\alpha_i)\; \forall i\in\{S,I,R\}$ for EGC MF model.

Thus, mean-field epidemiological models are more flexible tools than basic differential models, and more computationally simpler than agent-based ones. However, the need to correctly select the functional that describes the epidemiological situation, as well as the restrictions imposed on the model, lead to the need to solve complex and, in the general case, incorrect inverse problems.

\section{Acknowledgment}
The work was performed according to the Government research assignment for Sobolev Institute of Mathematics SB RAS, project FWNF-2024-0002.


\begin{thebibliography}{plain}

\bibitem{Lasry_2007} Lasry, J.-M. , Lions, P.-L.: Mean field games. Jpn. J. Math. 2 (1), 229–260 (2007) 

\bibitem{LionsLec2007} 
Lions, P.-L.: College de france course on mean-field games. Course of lectures. (2007--2011)

\bibitem{Huang2006} 
Huang, M., Malham$\acute{\rm e}$, R.P., Caines, P.E.: Large population stochastic dynamic games: closed-loop McKean-Vlasov systems and the Nash certainty equivalence principle. Commun. Inf. Syst. 6 (3), 221-251 (2006)

\bibitem{Huang2007}
Huang, M., Malham$\acute{\rm e}$, R.P., Caines, P.E. Large-population cost-coupled LQG problems with nonuniform agents: individual-mass behavior
and decentralized $\varepsilon $-Nash equilibria. IEEE Trans. Automat. Control. 52 (9), 1560-1571 (2006)

\bibitem{Roy_2023_EpidMFGOverview}  Roy, A., Singh, C., Narahari, Y.: Recent advances in modeling and control of epidemics using a mean field approach. Sådhanå. 48, 207 (2023)

\bibitem{Krivorotko_Kabanikhin_overview} Krivorotko, O.I., Kabanikhin, S.I.:	About mathematical modeling of COVID-19. Siberian Electronic Mathematical Reports. 20(2), p. 1211–1268 (2023)

\bibitem{Doncel_2020} Doncel, J., Gast, N., Gaujal, B.: A Mean Field Game Analysis of SIR Dynamics with Vaccination. Probability in the Engineering and Informational Sciences. 36(2), 1-18.(2020)  


\bibitem{Bremaud_2022}	Bremaud, L., Ullmo, D.: Social structure description of epidemic propagation with a mean-field game paradigm. Phys. Rev. E. 106, L062301 (2022)

\bibitem{Petr_SIRC_22}	Petrakova, V., Krivorotko, O.: Mean field game for modeling of covid-19 spread. Journal of Mathematical Analysis and Applications. 514 (1), 126271 (2022) 

\bibitem{Petrakova_IEEE2}
Petrakova, V., Krivorotko, O., Neverov, A.: Review of the mean field models for predicting the spread of viral infections.  IEEE Ural-Siberian Conference on Computational Technologies in Cognitive Science, Genomics and Biomedicine, 10329859 (2023)

\bibitem{Shaydurov_2020} Shaydurov, V., Kornienko, V., Zhang, S.: The Euler–Lagrange Approximation of the Mean Field Game for the Planning Problem. Lobachevskii J.Math. 41, 2703–2714 (2020)

\bibitem{Shaydurov_2021} Shaydurov, V., Kornienko, V.: Numerical Solution of Mean Field Problem with Limited Management Resource. Lobachevskii J.Math. 42, P. 1686–1696 (2021)

\bibitem{Kermack_27} Kermack, W.O., McKendrick, A.G.: A contribution to the mathematical theory of epidemics. Proceedings of the royal society of London. Series A. 115(772), 700-721 (1927)

\bibitem{Lee_21} Lee, W., Liu, S., Tembine, H., Li, W., Osher, S.:  Controlling Propagation of epidemics via mean-field control. SIAM J. Appl. Math. 81(1), 190-207 (2021)

\bibitem{Ben_2013} Bensoussan, A., Frehse, J.,  Yam, Ph.: Mean Field Games
	and Mean Field Type Control Theory. Springer, New York (2013)

\bibitem{Semendyaeva_22}
Semendyaeva, N.L., Orlov, M.V., Rui, T., Yang E.: Analytical and numerical investigation of the SIR mathematical model. Computational Mathematics and Modeling. 33(3), 284–299 (2022)

\bibitem{Aurell_Stackelberg} Aurell, A., Carmona, R., Dayanıklı, G., Lauriere, M.: Optimal incentives to mitigate epidemics: A Stackelberg mean field game approach. SIAM Journal on Control and Optimization. 60(2), 294-322 (2022).

\bibitem{Ullah_MFGFrac}
Ullah, M.F., Higazy, M., Ariful Kabir, K.M.: Dynamic analysis of mean-field and fractional-order epidemic vaccination strategies by evolutionary game approach. Chaos, Solitons \& Fractals. 162, 112431 (2022).


\bibitem{Tembine_MFG} Tembine, H.: Covid-19: data-driven mean-field-type game perspective. Games. 11(4), 51 (2020).

\bibitem{Gomes_ExistStationaryMFG}
Gomes, D.A., Patrizi, S., Voskanyan, V.: On the existence of classical solutions for stationary extended mean field games. Nonlinear Analysis: Theory, Methods \& Applications. 99, 49-79 (2014).

\bibitem{Achdou_MFG_monograph}
Achdou, Y., Cardaliaguet, P., Delarue, F., Porretta, A., Santambrogio, F.: Mean Field Games. Lecture Notes in Mathematics, 2281 (2019)

\bibitem{Bardi_NonUniq}
Bardi M., Fischer M.: On non-uniqueness and uniqueness of solutions in finite-horizon mean field games. ESAIM: Control, Optimisation and Calculus of Variations. 25, 44 (2019)

\bibitem{Briani_NonUniq}
Briani, A., Cardaliaguet, P.: Stable solutions in potential mean field game systems. Nonlinear Differential Equations and Applications. 25, 1-26  (2018)

\bibitem{Cecchin_NonUniq} Cecchin, A. et al.: On the convergence problem in mean field games: a two state model without uniqueness. SIAM Journal on Control and Optimization. 57(4), 2443-2466 (2019).

\bibitem{Poreta_2015}  Porretta, A.: Weak Solutions to Fokker–Planck Equations
and Mean Field Games. Arch. Rational Mech. Anal. 216, 1-62  1–62. (2015) 

\bibitem{Ataei_2023} Ataei, A.: Existence and uniqueness of the solutions to convection-diffusion equations. arXiv preprint arXiv:2310.20269.  (2023).


\bibitem{Krivorotko_2020} Krivorot’ko, O.I. et al.: Mathematical modeling and forecasting of COVID-19 in Moscow and Novosibirsk region. Numerical Analysis and Applications. 13, 332-348 (2020).


\bibitem{Saltelli_1999} Saltelli, A., Tarantola, S., Chan, K.P.S.: A quantitative model-independent method for global sensitivity analysis of model output. Technometrics. 41(1), 39-56 (1999).

\bibitem{Sobol_1990} Sobol, I.M.: On assessing the sensitivity of nonlinear mathematical models. Mathematical modeling. 2(1), 112-118 (1990) [In russian]

\bibitem{Petr_Sens_23}	Petrakova, V., Krivorotko, O.: Sensitivity of MFG SEIR-HCD Epidemiological Model. Lobachevskii Journal of Mathematics. 44(7),  2856-2869 (2023)

\bibitem{Schaibly_1973} Schaibly, J.H., Shuler, K.E.: Study of the sensitivity of coupled reaction systems to uncertainties in rate coefficients. II Applications. The Journal of Chemical Physics. 59(8), 3879-3888 (1973)

\bibitem{Krivorotko_SEIRHCD_2022} Krivorotko, O.I., Zyatkov, N.Y.: Data-driven regularization of inverse problem for SEIR-HCD model of COVID-19 propagation in Novosibirsk region. Eurasian Journal of Mathematical and Computer Applications. 10(1), 51-68 (2022)

\bibitem{Krivorotko_AgentModel_2021} Krivorotko, O. et al.: Agent-based modeling of COVID-19 outbreaks for New York state and UK: Parameter identification algorithm. Infectious Disease Modelling. 7(1), 30-44 (2022)





\end{thebibliography}
\end{document}